\documentclass[runningheads]{llncs}
\usepackage{graphicx}
\usepackage{tikz}
\usepackage{comment}
\usepackage{amsmath,amssymb} 
\usepackage{color}
\usepackage{booktabs}
\usepackage{multirow}
\usepackage{tabularx}
\usepackage{adjustbox}
\usepackage{wrapfig}
\usepackage{colortbl}
\usepackage[rightcaption]{sidecap}

\usepackage[accsupp]{axessibility}  
\usepackage{float}

\newcommand{\eat}[1]{}

\newcommand{\etal}{\textit{et al.}}

\usepackage{xcolor}
\usepackage{hyperref}
\hypersetup{
	colorlinks,
	linkcolor={red},
	citecolor={green},
	urlcolor={black}
}

\begin{document}
\pagestyle{headings}
\mainmatter
\def\ECCVSubNumber{1331}  

\title{LEDNet: Joint Low-light Enhancement and Deblurring in the Dark} 


\titlerunning{LEDNet}
%
\author{Shangchen Zhou \and Chongyi Li \and Chen Change Loy}
\authorrunning{S. Zhou et al.}
%
\institute{S-Lab, Nanyang Technological University, Singapore\\
    {\tt\small \{s200094, chongyi.li, ccloy\}@ntu.edu.sg}\\
    {\tt\small \url{https://shangchenzhou.com/projects/LEDNet}}}
\maketitle

\begin{abstract}
    Night photography typically suffers from both low light and blurring issues due to the dim environment and the common use of long exposure.
    While existing light enhancement and deblurring methods could deal with each problem individually, a cascade of such methods cannot work harmoniously to cope well with joint degradation of visibility and sharpness.
    Training an end-to-end network is also infeasible as no paired data is available to characterize the coexistence of low light and blurs.
    We address the problem by introducing a novel data synthesis pipeline that models realistic low-light blurring degradations,
    especially for blurs in saturated regions, e.g., light streaks, that often appear in the night images.
    With the pipeline, we present the first large-scale dataset for joint low-light enhancement and deblurring. The dataset, \textbf{LOL-Blur}, contains 12,000 low-blur/normal-sharp pairs with diverse darkness and blurs in different scenarios.
    We further present an effective network, named \textbf{LEDNet}, to perform joint low-light enhancement and deblurring.
    Our network is unique as it is specially designed to consider the synergy between the two inter-connected tasks.
    Both the proposed dataset and network provide a foundation for this challenging joint task.
    Extensive experiments demonstrate the effectiveness of our method on both synthetic and real-world datasets.
\end{abstract}

%

\section{Introduction}
\label{sec:intro}
\begin{figure}[thbp]
	\begin{center}
		\includegraphics[width=0.99\linewidth]{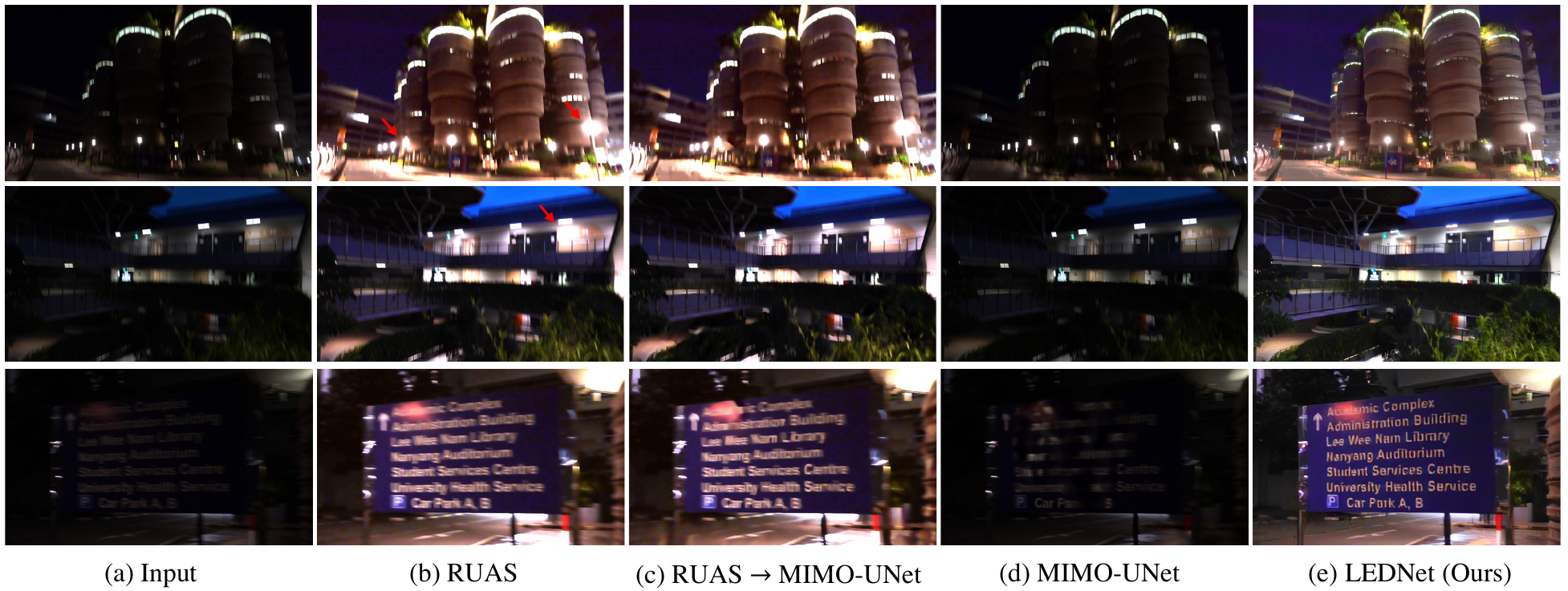}
		\caption{A comparison on the real-world night blurry images shows that existing low-light enhancement and deblurring methods fail in coping with the night blurry images.
				(a) Input images.
				(b) Motion blur in the saturated area is enlarged after performing light enhancement using a contemporary method RUAS~\cite{liu2021retinex} (indicated by red arrows).
				%
				(c) Applying the deblurring network MIMO-UNet~\cite{cho2021rethinking} after light enhancement still fails in blur removal.
				%
				%
				(d) MIMO-UNet trained on daytime GoPro dataset fails to remove blur in the nighttime images.
				(e) The proposed LEDNet trained with our  LOL-Blur dataset yields satisfactory results through joint low-light enhancement and deblurring.
			}
		\label{fig:teaser}
	\end{center}
\end{figure}
\begin{figure}[t]
	\begin{center}
		\includegraphics[width=\linewidth]{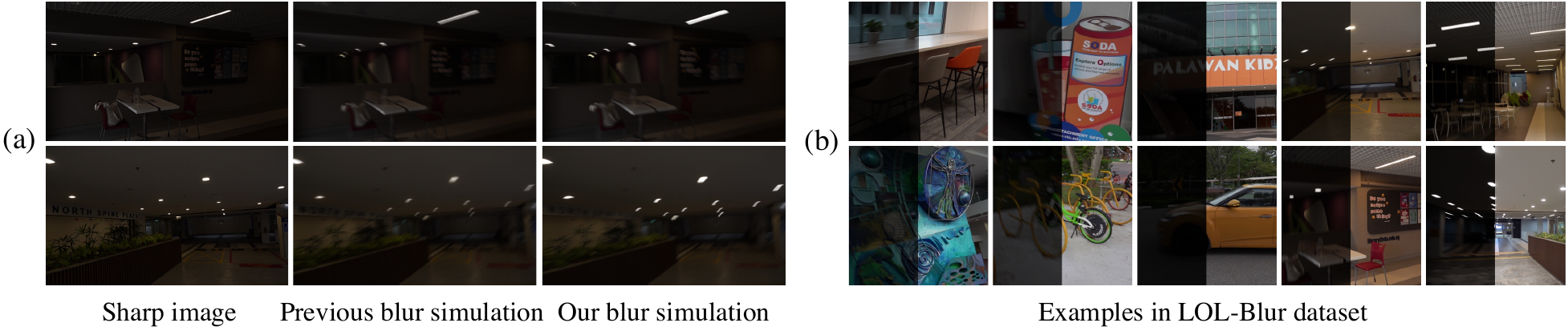}
		\caption{
				(a) Comparison on blur simulation. The previous blur simulation~\cite{su2017deep,nah2017deep,Nah2019REDS}, i.e., simple averaging of a sharp sequence, tends to weaken blurs in  saturated pixels. By contrast, our simulation generates more realistic saturated blurs that maintain the saturated intensities during blur synthesis.
				(b) Paired images in the LOL-Blur dataset, containing diverse darkness and blurs (saturated and unsaturated)  in the dark.
				%
			}
		\label{fig:intro_blur_data}
	\end{center}
\end{figure}

When capturing images at night, one would usually use a slow shutter speed (long exposure) to allow more available light to illuminate the image.
%
%
%
Even so, the captured dark images may still suffer from low visibility and distorted color induced by insufficient light, which is constrained by minimum shutter speeds that are acceptable for handheld shooting in the dark.
Annoyingly, long exposure inevitably causes motion blurs due to camera shake and dynamic scenes.
%
Thus, both low light and motion blurs typically co-exist in in the night images.
Prior methods address the two tasks independently, i.e., low-light enhancement~\cite{LoLi2021,wei2018deep,zerodce} and image deblurring~\cite{hu2014deblurring,nah2017deep,tao2018scale,zhang2019deep,kupyn2019deblurgan,zhang2020deblurring,chen2021blind}.
These methods made independent assumptions in their specific problem. As a result, a forceful combination cannot solve the joint degradation caused by low light and motion blur.
%
%
%
Specifically, existing low-light enhancement methods~\cite{wei2018deep,liu2021retinex} perform intensity boosting and denoising, ignoring spatial degradation of motion blurs.
Instead, motion blur is even enlarged in saturated regions due to over-exposing after performing light enhancement, as shown in Fig.~\ref{fig:teaser}(b).
Low-light enhancement methods~\cite{wei2018deep,zhang2019kindling} also have the risk of removing informative clues for blur removal due to over-smoothing while denoising.
Fig.~\ref{fig:teaser}(c) shows that performing deblurring after light enhancement fails the blur removal.
As for deblurring, existing methods~\cite{tao2018scale,zhang2019deep,kupyn2019deblurgan,cho2021rethinking} trained on the datasets that only contain daytime scenes,
%
and thus, cannot be directly applied to the non-trivial night image deblurring.
In particular, motion cues in dark regions are poorly visible and perceived due to the low dynamic range,
posing a great challenge for these existing deblurring methods.
Furthermore, night blurry images contain saturated regions (e.g., light streaks) in which the pixels do not conform to the blur model learned from daytime data~\cite{hu2014deblurring,chen2021blind}.
As observed in Fig.~\ref{fig:teaser}(d), the deblurring network trained on daytime GoPro dataset fails to remove blur in the night images.
The solution to the aforementioned problems is to train a single network that addresses both types of degradations jointly.
Clearly, the main obstacle is the availability of such data that come with low-light blurry and normal-light sharp image pairs. The collection is laborious and hard, if not impossible.
Existing datasets for low-light enhancement, e.g., LOL~\cite{wei2018deep} and SID~\cite{chen2018learning}, gather low-/normal-light pairs by changing exposure time and ISO in two shots.
While deblurring datasets, e.g., RealBlur~\cite{rim2020real}, need to capture paired blurry/sharp images under the long and short exposures using a dual-camera system.
%
However, it is challenging to merge these two data collection processes harmoniously to capture paired data for this joint task.
%
%
Moreover, the existing synthetic deblurring datasets~\cite{su2017deep,nah2017deep,Nah2019REDS} (e.g., GoPro) cannot simulate blurs of saturated regions for night images
due to the lack of sequences that contain saturated scenes and their inappropriate blur simulation by simply averaging.
%
%
As observed in Fig.~\ref{fig:intro_blur_data}(a) (middle), their blur simulation method tend to undesirably attenuate blurs of saturated areas, which do not resemble the real ones shown in Fig.~\ref{fig:teaser}(a).

This paper makes the first attempt to propose a novel data synthesis pipeline for joint low-light enhancement and deblurring.
%
In particular, we circumvent the difficulty of obtaining dark and blurry images through a heuristic approach for simulating low-light degradation,
and a new blur simulation method that pays special attention to model blurs in saturated regions correctly.
Figure~\ref{fig:intro_blur_data} shows that our blur simulation generates more realistic saturated blurs than the previous ones~\cite{su2017deep,nah2017deep,Nah2019REDS}, maintaining saturated intensities during blur synthesis.
%
%
%
%
The resulting dataset, \textbf{LOL-Blur}, contains 12,000 pairs of low-blur/normal-sharp pairs for training and testing.
Examples of LOL-Blur are shown in Fig.~\ref{fig:intro_blur_data}(b).

Apart from the data, we show that it is beneficial to consider both low-light enhancement and deblurring in a single context. In particular, we demonstrate a novel encoder-decoder pipeline, \textbf{L}ow-light \textbf{E}nhancement and \textbf{D}eblurring \textbf{Net}work (\textbf{LEDNet}), where the encoder is specialized in light enhancement and the decoder in deblurring. The encoder and decoder are linked with adaptive skip connections. This unique structure allows the passing of light enhanced features in the decoder for blur removal in the decoder.
The main contributions: 1) We introduce a novel data synthesis pipeline that models low-light blur degradation realistically, leading to the large-scale and diverse LOL-Blur dataset for joint low-light enhancement and deblurring. 2) We propose a unified network LEDNet with delicate designs to address low-light enhancement and deblurring jointly. 3) We present to aggregate hierarchical global prior that is crucial for stable training and artifacts suppression, as well as the learnable non-linear layer helps brighten dark areas without overexposing other regions. 4) Extensive experiments show that our method achieves superior results to prior arts on both synthetic and real-world datasets.
%


\section{Related Work}
\noindent {\bf Image Deblurring.}
Many CNN-based methods have been proposed for dynamic scene deblurring~\cite{nah2017deep,tao2018scale,zhang2018dynamic,zhang2019deep,kupyn2019deblurgan,zhang2020deblurring,cho2021rethinking}.
Most early studies~\cite{sun2015learning,gong2017motion} employ networks to estimate the motion blur kernels followed by non-blind methods.
Owing to the emergence of training datasets for deblurring tasks~\cite{su2017deep,nah2017deep,Nah2019REDS,shen2019human,zhou2019davanet,li2021arvo,rim2020real},
end-to-end kernel-free networks become the dominant methods.
To obtain a large receptive field, some networks~\cite{nah2017deep,tao2018scale,cho2021rethinking} adopt a multi-scale strategy to handle large blurs
%
Similarly, some multi-patch deblurring networks~\cite{zhang2020deblurring,zamir2021multi,Hu_2021_ICCV} employ the hierarchical structures without down-sampling.
GAN-based deblurring methods~\cite{kupyn2018deblurgan,kupyn2019deblurgan} have been proposed to generate more details.
To deal with spatially-varying blurs, Zhang \textit{et al.}~\cite{zhang2018dynamic} propose spatially variant RNNs to remove blur via estimating RNN weights.
%
%
Zhou \textit{et al.}~\cite{zhou2019spatio} propose the filter adaptive convolutional (FAC) layer to handle non-uniform blurs dynamically.
%
%
In our paper, we built a filter adaptive skip connection between encoder and decoder using FAC layers.
%

%
Optimization-based approaches are proposed for low-light image deblurring~\cite{hu2014deblurring,chen2020oid,chen2021blind,Chen_2021_CVPR}.
Hu \textit{et al.}~\cite{hu2014deblurring} suggest the use of light streaks to estimate blur kernel.
However, their method heavily relies on light streaks and tends to fail when the light sources are not available or too large beyond pre-designed blur kernel size.
Chen \textit{et al.}~\cite{chen2021blind,Chen_2021_CVPR} process saturated regions specially and ensure smaller contributions of these pixels in optimization.
Their results show few artifacts around saturated regions.
%
While effective, all these methods are time-consuming, thus limiting their applicability.

\noindent {\bf Low-light Enhancement.}
Deep networks have become the mainstream in low-light enhancement (LLE)~\cite{LoLi2021}.
The first CNN model LL-Net~\cite{lore2017llnet} employs an autoencoder to learn denoising and light enhancement simultaneously.
Inspired by the Retinex theory, several LLE networks~\cite{wei2018deep,zhang2019kindling,wang2019underexposed,yang2021sparse,liu2021retinex} are proposed.
They commonly split a low-light input into reflectance and illumination maps, then adjust the illumination map to enhance the intensity.
Most methods integrate a denoising module on the reflectance map for suppressing noise in the enhanced results.
For example, Zheng \textit{et al.}~\cite{zheng2021adaptive} propose an unfolding total variation network to estimate noise level for LLE.
%
While the joint task of LLE and deblurring has not been investigated yet in the literature.
%

%
%
To improve the generalization capability, some unsupervised methods are proposed.
EnlightenGAN~\cite{jiang2021enlightengan} is an attention-based U-Net trained using adversarial loss.
%
%
Zero-DCE~\cite{zerodce} and Zero-DCE++~\cite{zerodce++} formulate light enhancement as a task of image-specific curve estimation.
Their training adopts several manually-defined losses on supervision of exposure or color, without limitation of paired or unpaired training data.
Thus, Zero-DCE can be easily extended to generic lighting adjustments. Notably, due to the pixel-wise curve adjustment formulation, it can be used for spatially-varying light adjustment.
In our data synthesis pipeline, we train an exposure conditioned Zero-DCE to darken images for low-light simulation.
Given random low exposure degrees, we can generate low-light images of diverse darkness.
%
%
\begin{figure*}[t]
	\begin{center}
		\includegraphics[width=\linewidth]{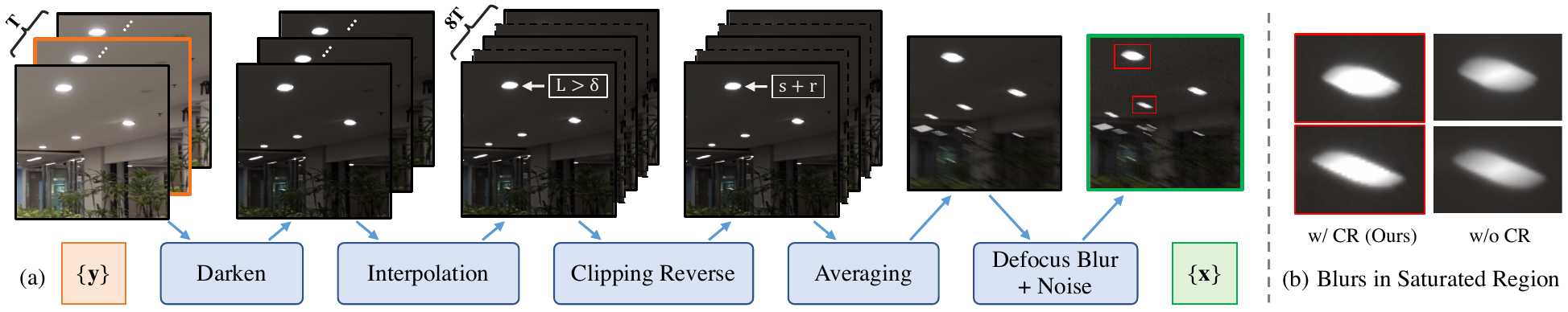}
		\caption{{(a) An overview of our data synthesis pipeline. (b) Comparisons on two blur simulations in the saturated regions.
				With the proposed Clipping Reverse (CR), we can generate realistic blurs with sharp boundaries in saturated regions, which better resembling  real cases that are caused by the large light ratio in night photography.}}
		\label{fig:data_pipeline}
	\end{center}
\end{figure*}
\section{LOL-Blur Dataset}
%

Efforts have been made to collect real-world paired data for low-light enhancement~\cite{wei2018deep,liu2021benchmarking,chen2018learning,chen2019seeing,jiang2019learning,wang2021seeing} or image deblurring~\cite{rim2020real}, but not both settings at the same time.
The lack of such data is not surprising as (1) Paired images of low-light enhancement datasets and image deblurring datasets are commonly collected by different camera shot settings, and
(2) The collection of both kinds of data is susceptible to geometric and photometric misalignment due to camera shake or dynamic environment during data acquisition.

In this work, inspired by the big success of data synthesis in the real-world super-resolution tasks~\cite{wang2021realesrgan,zhang2021designing,li2020blind,chan2021realbasicvsr}, we introduce a synthesis pipeline that models low-light blur degradation jointly, hence allowing us to generate a large-scale dataset (LOL-Blur).
We acquire a total of 170 videos for training and 30 videos for testing, each of which has 60 frames, amounting to 12,000 paired data.
%
\subsection{Existing Synthesis Methods and Limitations}
%
%
%
\noindent {\bf Low-light Simulation.} Prior works~\cite{lore2017llnet,lv2018mbllen} use Gamma correction to simulate low-light images, defined by a nonlinearity power-law expression $I_{low} = \alpha I_{in}^{\gamma}, (\gamma>1)$, where constant $\alpha$ is usually set to 1.
This synthetic process tends to introduce large color deviation with noticeable warm tones~\cite{LoLi2021}, in contrast, our simulation method EC-Zero-DCE (refer to Sec.~\ref{sec: data_pipeline}) produce more natural and realistic low-light images.
A comparison is provided in the supplementary.
\noindent {\bf Blur Simulation.} A standard synthesis pipeline of blurry data~\cite{su2017deep,nah2017deep,Nah2019REDS,zhou2019davanet,li2021arvo} is to average successive frames on high frame-rate sequences for approximating the blur model~\cite{nah2017deep}.
The process can be expressed as:
\begin{equation}\label{eq: blur}
	\resizebox{0.6\linewidth}{!} {$
			B = g\left(\frac{1}{T} \sum_{i=0}^{T-1} S[i] \right) = g\left(\frac{1}{T} \sum_{i=0}^{T-1} g^{-1}\left(\hat{S[i]}\right) \right),
		$}
\end{equation}
where $g(\cdot)$ is CRF function (Gamma curve with $\gamma = 2.2$) that maps latent signal $S[i]$ into observed sRGB images $\hat{S[i]}$.
This process can be used to generate blurry-sharp pairs for daytime scenes, assuming $\hat{S[i]} = g\left(S[i]\right)$.
This blur model, however, is usually not accurate to the regions of saturated pixels that often appear in dark blurry images.
This is because the saturated intensities in latent signal $S[i]$ are clipped to the maximum value (255) when $S[i]$ is saved as an sRGB image $\hat{S[i]}$, due to the limited dynamic range of sRGB images, i.e., $\hat{S[i]} = Clip\left(g\left(S[i]\right)\right)$.
This clipping function damages the exceeding value of saturated regions, thus making the blur model of Eq.~\eqref{eq: blur} improper for these regions~\cite{chen2021blind}.
Our simulation pipeline resolves this issue by recovering the clipped intensities in saturated regions (refer to Sec.~\ref{sec: data_pipeline}), generating more realistic light blurs.
As a visual comparison are shown in Fig.~\ref{fig:intro_blur_data}(a).
\subsection{Data Generation Pipeline}
\label{sec: data_pipeline}
The overview of our data generation pipeline is shown in Fig.~\ref{fig:data_pipeline}.
We use a Sony RX10 IV camera to record 200 high frame-rate videos at 250 fps.
%
%
With the video sequences, we first downsize each frame to a resolution of $1120 \times 640$ to reduce noises.
We then apply VBM4D~\cite{maggioni2012video} for further denoising and obtain the clean sequences.
In our method, we take 7 or 9 frames as a sequence clip, as shown in Fig.~\ref{fig:data_pipeline}.
The mid-frame (with orange bounding box) among the sharp frames is treated as the ground truth image.
Then, we process the following steps to generate low-light and blurred images.
\noindent {\bf Darkening with Conditional Zero-DCE.}
To simulate the degradation of low light, we reformulate the Zero-DCE~\cite{zerodce} into an Exposure-Conditioned variant, EC-Zero-DCE.
Contrary to Zero-DCE that is designed for improving the brightness of an image, EC-Zero-DCE simulates low light with controllable darkness levels, via implementing a reversed curve adjustment.
Specifically, we modify the exposure control loss by replacing the fixed exposure value with a random parameter that represents darkness while other losses are kept in the same settings as Zero-DCE.
Given different exposure levels, EC-Zero-DCE can generate realistic low-light images with diverse darkness.
Note that EC-Zero-DCE performs pixel-wise and spatially-varying light adjustment, rather than uniform light degradation. We provide the luminance adjustment map in the supplementary to support this statement.
\noindent {\bf Frame Interpolation.}
To avoid discontinuous blurs in the synthetic blurry images, we increase the frame rate to 2000 fps using a high-quality frame interpolation network~\cite{niklaus2017video}.
\noindent {\bf Clipping Reverse for Saturated Region.}
To restore the clipped intensity in saturated regions that were ignored by the previous blur simulation (i.e., Eq.~\eqref{eq: blur}),
a simple yet effective way is by adding a  random supplementary value $r\sim \mathcal U(20, 100)$ to RGB channels in these regions.
We first define the saturated regions where lightness channel $L > \delta$ in the Lab color space,  the threshold $\delta$ is empirically set to 98 in our pipeline, where $L \in [0, 100]$.
Then, we reformulate the blur model as a more general form for both saturated and unsaturated regions, as shown in Eq.~\eqref{eq: blur_new}:
\begin{equation}\label{eq: blur_new}
	\resizebox{0.45\linewidth}{!} {$
			B =  g\left(\frac{1}{T} \sum_{i=0}^{T-1} Clip^{-1}\left(g^{-1}\left(\hat{S[i]}\right) \right)\right),
		$}
\end{equation}
where $Clip^{-1}(s) = s + r$ if $s$ in the saturated regions, otherwise $Clip^{-1}(s) = s$.
Fig.~\ref{fig:intro_blur_data}(a) and Fig.~\ref{fig:data_pipeline}(b) shows our blur simulation using clipping reverse (w/ CR) generates more realistic saturated blurs than the GoPro blur simulation (w/o CR) that is commonly used in previous datasets.
Moreover, the modified blur simulation (w/ CR) indeed helps networks handle well on both unsaturated and saturated blurs, as indicated by the comparison in Fig.~\ref{fig:blur_simulation}.
\noindent {\bf Frame Averaging.}
Next, we average 56 ($7\times 8$) or 72 ($9\times 8$) successive frames of 2000 fps videos to produce virtual blurry videos at  around 24 fps.
%

%
\noindent {\bf Adding Defocus Blur and Noise.}
To generate more realistic low-light blurry images, our pipeline also considers defocus blurs by applying generalized Gaussian filters~\cite{wang2021realesrgan}.
We also add realistic noises into low-blur images generated by CycleISP~\cite{zamir2020cycleisp}.
Both defocus blur and noise are added in a random fashion.

\noindent {\bf Discussion.}
Our dataset offers realism in low-light blur degradation and consists of 200 common dynamic dark scenarios (indoor and outdoor) with diverse darkness and motion blurs, as shown in Fig.~\ref{fig:intro_blur_data}(b).
Compared to previous synthetic deblurring datasets (such as GoPro~\cite{nah2017deep} and REDS~\cite{Nah2019REDS}) that only contain daytime scenes and lacks the saturated regions, our dataset contains a total of 55 sequences with various sources of artificial lights that often appear in the night photography. Hence, our simulated data sufficiently covers hard cases with blurs in saturated areas, e.g., light streaks, which are indispensable for our joint task.
Experimental results demonstrate that the networks trained using our dataset generalizes well on real-world dark blurred images.
%
%
%
\section{LEDNet}
\begin{figure*}[t]
	\centering
	\includegraphics[width=0.98\textwidth]{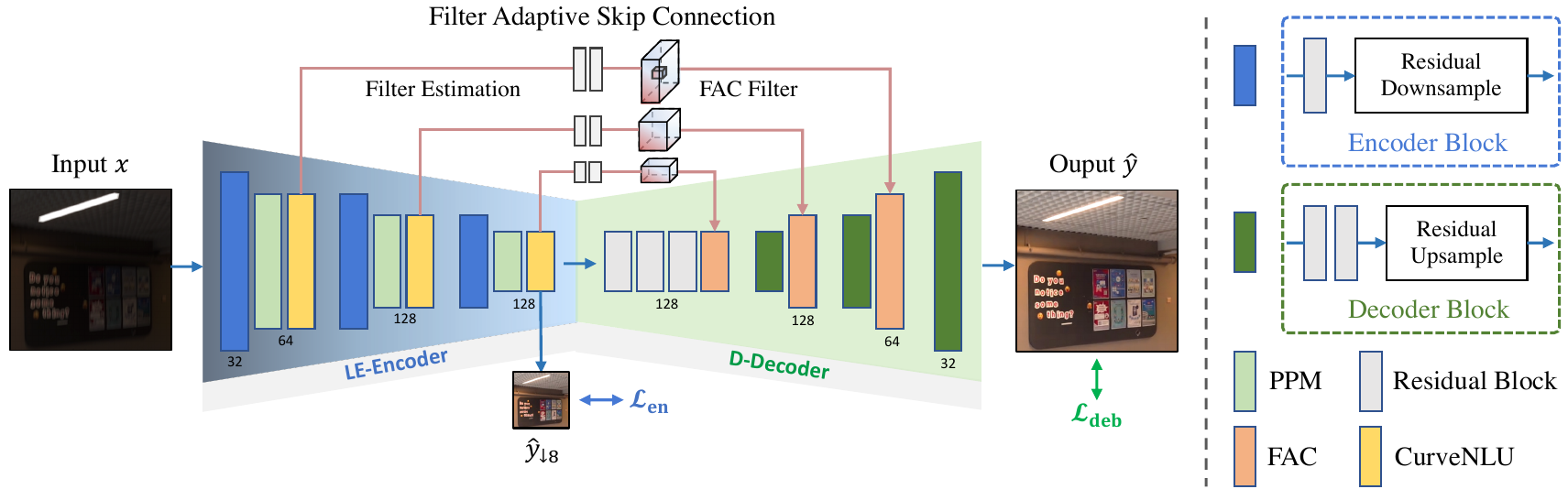}
	\caption{{An illustration of the proposed LEDNet. It contains an Encoder for Light Enhancement, LE-Encoder, and a Decoder for Deblurring, D-Decoder. They are connected by three Filter Adaptive Skip Connections.
			The PPM and CurveNLU layers are inserted in LE-Encoder, making light enhancement more stable and powerful.
			LEDNet applies spatially-adaptive transformation to D-Decoder using filters generated by FASC from enhanced features.
			CurveNLU and FASC enable LEDNet to perform spatially-varying feature transformation for both intensity enhancement and blur removal.
		}}
	\label{fig:pipeline}
\end{figure*}

We treat the joint task of low-light enhancement (LLE) and deblurring as a non-blind image restoration problem.
The low-light blurry images $\{x\}$ contain the mixed degradations of visibility and texture.
The two type degradations are spatially-varying due to local lighting conditions and dynamic scene blurs.
To solve this issue, we specially design a network, LEDNet, to map low-light blurry images $\{x\}$ to its corresponding normal-light sharp images $\{y\}$.
As shown in Fig.~\ref{fig:pipeline}, LEDNet is built upon an encoder-decoder architecture with filter adaptive skip connections to solve this joint spatially-varying task.
\subsection{Low-light Enhancement Encoder}
The encoder (LE-Encoder) is designed for Low-light Enhancement with the supervision of intermediate enhancement loss (see Sec.~\ref{sec:loss}).
It consists of three scale blocks, each of which contains one Residual Block, one Residual Downsampling Block~\cite{zamir2020learning}, a Pyramid Pooling Module (PPM)~\cite{zhao2017pyramid}, and a Curve Non-Linear Unit (CurveNLU), as shown in Fig.~\ref{fig:pipeline}.
To facilitate intermediate supervision, we output an enhanced image by one convolution layer at the smallest scale.
Our design gears LE-Encoder to embed the input image $x$ into the feature space of normal-light images, allowing the subsequent decoder (D-Decoder) to pay more attention to the deblurring task.
\noindent {\bf Pyramid Pooling Module.}
%
%
The outputs of typical light enhancement networks are often prone to local artifacts, especially when the networks are fed with high-resolution inputs.
We found that the problem can be significantly remedied by injecting global contextual prior into the networks.
To achieve this goal, we introduce PPM into our LE-Encoder.
The PPM extracts hierarchical global prior using multi-scale regional pooling layers and aggregates them in the last convolution layer.
We adopt the original design of PPM that has four mean pooling branches with bin sizes of $1, 2, 3, 6$, respectively.
The is the first time PPM is used in a low-light enhancement network. We show that it is crucial for suppressing artifacts that may be caused by the co-existence of other degradations of blur and noise (refer to a comparison shown in Fig.~\ref{fig:ppm}).
%
%

%
\noindent {\bf Curve Non-Linear Unit.}
Local lighting such as light sources are often observed in the night environment.
A global operator tends to over- or under-enhance these local regions.
To solve this problem, Zero-DCE~\cite{zerodce} applies pixel-wise curve parameters to the input image iteratively for light enhancement.

\begin{SCfigure}[1][t]
	\includegraphics[width=0.5\textwidth]{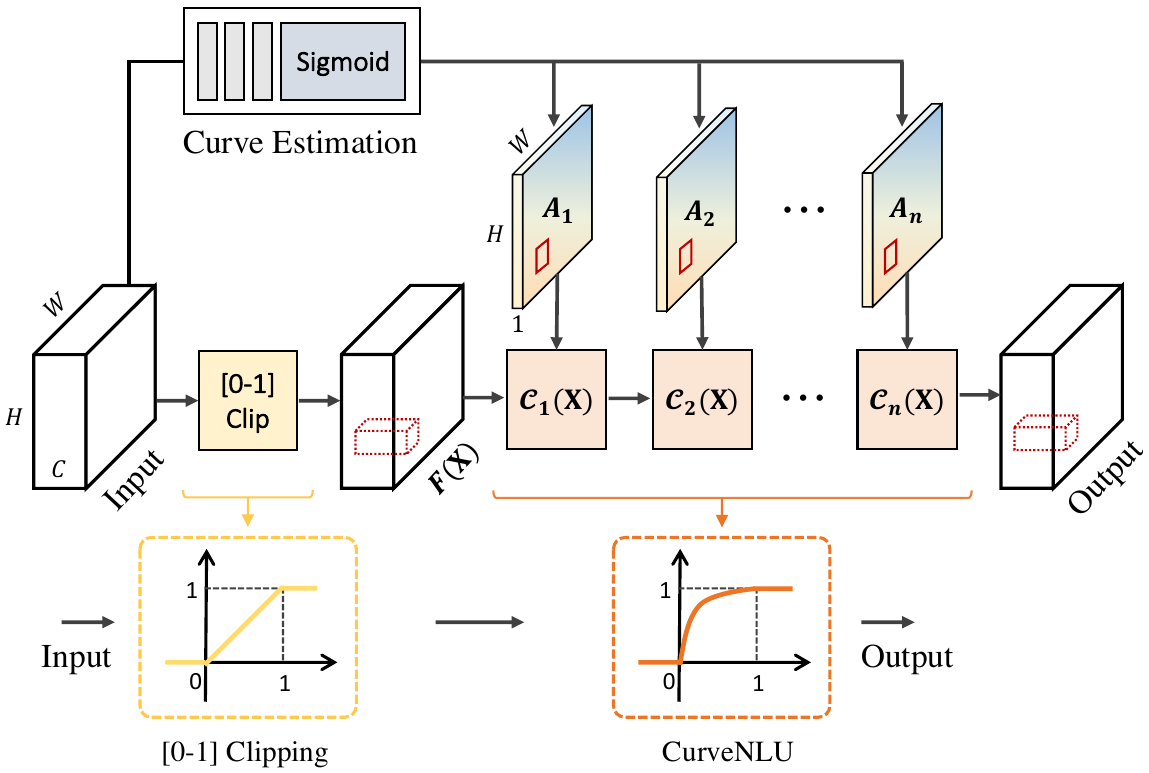}
	\caption{{An illustration of Curve Non-Linear Unit.
			This layer can be seen as a learnable non-linear activation function between 0 and 1.
			Based on Eq.~\ref{eq: curvenlu}, the learned function always follows concave down increasing curves to increase feature intensities.
		}}
	\label{fig:curvenlu}
\end{SCfigure}

Inspired by Zero-DCE, we propose a learnable non-linear activation function, namely CurveNLU.
The CurveNLU is designed for feature transformation using the estimated curve parameters, as shown in Fig.~\ref{fig:curvenlu}.
Similar to Zero-DCE, we formulate the high-order curve in an iterative function:
\begin{equation}\label{eq: curvenlu}
	\resizebox{0.7\linewidth}{!} {$
			\mathcal{C}_n(\mathrm{p}) = \left\{ \begin{array}{ll}
				A_1 F(\mathrm{p})(1-F(\mathrm{p})) + F(\mathrm{p}),                                                                & n = 1 \\
				\vspace{-2mm}
				\\
				A_{n-1}(\mathrm{p})\mathcal{C}_{n-1}(\mathrm{p})(1-\mathcal{C}_{n-1}(\mathrm{p})) + \mathcal{C}_{n-1}(\mathrm{p}), & n > 1
			\end{array} \right.
		$
	}
\end{equation}
where $\mathrm{p}$ denotes position coordinates of features, and $A_{n-1}$ is the pixel-wise curve parameter for the n-$th$ order of the estimated curve.
Given an input feature $F\in \mathbb{R}^{H\times W\times C}$, Curve Estimation module estimates curve parameters $A \in \mathbb{R}^{H\times W\times n}$ that represent an $n+1$ order curve for different positions.
Feature transformation is then achieved by Eq.~\ref{eq: curvenlu} using the estimated curve parameters.
Different from Zero-DCE that uses different curves for RGB channels, our CurveNLU applies the same curve to different channels in the feature domain.
Note that the parameters $A$ lay in $\left[0, 1\right]$, ensuring that CurveNLU always learns concave down increasing curves to increase the features of dark areas without overexposing other regions.
To meet such a requirement, the input feature $F$ of CurveNLU is needed to be clipped to the range of $[0, 1]$ at the beginning.
The Curve Estimation module consists of three convolution layers followed by a Sigmoid function.
We set the iteration number $n$ to 3 in our experiments.
%
%
\subsection{Deblurring Decoder}
%
With the enhanced features from LE-Encoder, Deblurring Decoder (D-Decoder) is able to concentrate more on deblurring.
It contains three convolutional blocks, each of which has two Residual Blocks, one Residual Upsampling Block~\cite{zamir2020learning},
and a FAC Layer~\cite{zhou2019spatio} that is used to bridge the LE-Encoder and the D-Decoder.
\subsection{Filter Adaptive Skip Connection}
Both low-light enhancement and deblurring in our task are spatially varying problems.
Deblurring in the dynamic scenes is challenging due to its spatially variant blurs caused by object motion and depth variations.
Though CurveNLU applies pixel-wise adjustment in the LE-Encoder, it is not enough for the deblurring task that usually needs dynamic spatial kernels to handle motion blurs.
Filter Adaptive Convolutional (FAC) layer \cite{zhou2019spatio} has been proposed to apply dynamic convolution filters for each element in features.
Built on the FAC layers, we design a Filter Adaptive Skip Connection (FASC) to solve the deblurring problem by exploiting the enhanced information from LE-Encoder.
As shown in Fig.~\ref{fig:pipeline}, given the enhanced features $E\in \mathbb{R}^{H\times W\times C}$ at different scales, FASC estimates the corresponding filter $K\in \mathbb{R}^{H\times W\times Cd^2}$ via three $3\times 3$ convolution layers and a $1\times 1$ convolution layer to expand the feature dimension.
The filter $K$ is then used by FAC layers to transform the features $D \in \mathbb{R}^{H\times W\times C}$ in D-Decoder.
For each element of feature $D$ , FAC applies a convolution operator using the corresponding $d\times d$ kernel from the filter $K$ to obtain refined features.
We set the kernel size $d$ to 5 at the three scales, following the setting in Zhou \textit{et al}.~\cite{zhou2019spatio}.
%
%
\subsection{Loss Function}
\label{sec:loss}
\noindent {\bf Low-light Enhancement Losses.}
To provide intermediate supervision, we employ L1 loss and perceptual loss at $\times 8$ downsampled scale.
Specifically, we predict the image $\hat{y}_{\downarrow 8}$ for the smallest scale of LE-Encoder, and then restrict it using scale-corresponding ground truth $y_{\downarrow 8}$, shown as Eq.~\eqref{eq:enhance_loss}:
\begin{align}
	\label{eq:enhance_loss}
	\mathcal{L}_{en} =\left\| \hat{y}_{\downarrow 8} - y_{\downarrow 8}\right\|_1 + \lambda_{per}\left\| \phi\left(\hat{y}_{\downarrow 8}\right) - \phi\left(y_{\downarrow 8}\right)\right\|_1,
\end{align}
where $\phi (\cdot)$ represents the pretrained VGG19 network. We adopt multi-scale feature maps from layer $\left\{conv1, \cdots, conv4 \right\}$ following the widely-used setting~\cite{wang2018esrgan}.
Due to downsampling space, the enhancement loss $\mathcal{L}_{en}$ mainly supervises the exposure of intermediate output.
%
%

%
\noindent {\bf Deblurring Losses.}
We use the L1 loss and perceptual loss as our deblurring loss $\mathcal{L}_{deb}$, defined as follows:
\begin{align}
	\label{eq:deblur_loss}
	\mathcal{L}_{deb} =\left\| \hat{y} - y \right\|_1 + \lambda_{per}\left\| \phi\left(\hat{y}\right) - \phi\left(y\right)\right\|_1.
\end{align}
The overall loss function is:
\begin{align}
	\label{eq:overall_loss}
	\mathcal{L} =\lambda_{en}\mathcal{L}_{en} + \lambda_{deb}\mathcal{L}_{deb}.
\end{align}
%
We set the loss weights of $\lambda_{per}$, $\lambda_{en}$, and $\lambda_{deb}$ to 0.01, 0.8, and 1, respectively.

\section{Experiments}
\label{sec:exp}
%
%
\noindent\textbf{Dataset and Experimental Settings.}
We train our network LEDNet and other baselines on LOL-Blur dataset.
The 170 sequences (10,200 pairs) are used for training and 30 sequences (1,800 pairs) for test.
We randomly crop $256\times 256$ patches for training.
%
%
The mini-batch size is set to 8.
We train our network using Adam optimizer with $\beta_1 = 0.9, \beta_2 = 0.99$ for a total of 500k iterations.
The initial learning rate is set to $10^{-4}$ and updated with cosine annealing strategy~\cite{loshchilov2016sgdr}.
%
%
%
%
%

%
\subsection{Evaluation on LOL-Blur Dataset}
\label{sec:lolblur_evaluation}
We quantitatively and qualitatively evaluate the proposed LEDNet on our LOL-Blur Dataset.
Since the joint task is newly-defined in this paper, there is no method available to make a comparison directly.
We carefully choose and combine existing representative low-light enhancement and deblurring methods, providing three types of baselines for comparisons.
%
%
%
%
%
Specifically, the baseline methods lay on following three categories:

\noindent
\textbf{1. Enhancement $\rightarrow$ Deblurring.}
We choose the recent representative light enhancement networks Zero-DCE~\cite{zerodce} and RUAS~\cite{liu2021retinex} followed by a state-of-the-art deblurring network MIMO-UNet~\cite{cho2021rethinking}.

\noindent
\textbf{2. Deblurring $\rightarrow$ Enhancement.}
For deblurring, we include a recent optimization-based method~\cite{chen2021blind} particularly designed for saturated image deblurring, a GAN-based network DeblurGAN-v2~\cite{kupyn2019deblurgan} trained on RealBlur dataset, and a state-of-the-art deblurring network MIMO-UNet~\cite{cho2021rethinking} trained on GoPro dataset.
Since RUAS tends to produce overexposed results in the saturated regions that may cover up previous deblurring results, we employ Zero-DCE for light enhancement in this type of baseline.
%

\noindent
\textbf{3. End-to-end training on LOL-Blur dataset.}
We retrain some state-of-the-art baselines on our dataset using their released code. They include two light enhancement networks KinD++~\cite{zhang2021beyond} and DRBN~\cite{yang2020fidelity}, and three deblurring networks of DeblurGAN-v2~\cite{kupyn2019deblurgan}, DMPHN~\cite{zhang2019deep}, and MIMO-UNet~\cite{cho2021rethinking}.
\begin{table*}[t]
	\centering
	\caption{{Quantitative evaluation on our LOL-Blur dataset. PSNR/SSIM$\uparrow$: the higher, the better; LPIPS$\downarrow$: the lower, the better.
			The symbol `$^\dagger$' indicates that we use DeblurGAN-v2 trained on RealBlur dataset, and `$^*$' indicates the network is retrained on our LOL-Blur dataset.
			The runtime and parameters are expressed in seconds and millions.
			All runtimes are evaluated on a 720p image using a  GPU V100.
		}}
	\resizebox{\linewidth}{!} {
		\begin{tabular}{lccc|cccc|cccccc}
			\toprule
			\multirow{3}{*}{Methods} & \multicolumn{2}{c}{Enhancement $\rightarrow$ Deblurring} &
			                         & \multicolumn{3}{c}{Deblurring $\rightarrow$ Enhancement} &
			                         & \multicolumn{5}{c}{Training on LOL-Blur}                                                                                                                                                                                                                                 \\
			\cline{2-4}     \cline{5-8}   \cline{9-14} 
			                         & Zero-DCE~\cite{zerodce}                                  & RUAS~\cite{liu2021retinex}                       &
			                         & Chen~\cite{chen2021blind}                                & DeblurGAN-v2$^\dagger$~\cite{kupyn2019deblurgan} & MIMO~\cite{cho2021rethinking}          &
			                         & KinD++$^*$                                               & DRBN$^*$                                         & DeblurGAN-v2$^*$                       & DMPHN$^*$             & MIMO$^*$                  & Ours                                                          \\
			                         & $\rightarrow$ MIMO~\cite{cho2021rethinking}              & $\rightarrow$ MIMO~\cite{cho2021rethinking}      &
			                         & $\rightarrow$ Zero-DCE~\cite{zerodce}                    & $\rightarrow$  Zero-DCE~\cite{zerodce}           & $\rightarrow$  Zero-DCE~\cite{zerodce} &
			                         & ~\cite{zhang2021beyond}                                  & ~\cite{yang2020fidelity}                         & ~\cite{kupyn2019deblurgan}             & ~\cite{zhang2019deep} & ~\cite{cho2021rethinking} &                                                               \\
			\midrule
			PSNR$\uparrow$           & 17.68                                                    & 17.81                                            &                                        & 17.02                 & 18.33                     & 17.52 &  & 21.26 & 21.78 & 22.30 & 22.20 & 22.41 & \bf{25.74} \\
			SSIM$\uparrow$           & 0.542                                                    & 0.569                                            &                                        & 0.502                 & 0.589                     & 0.57  &  & 0.753 & 0.768 & 0.745 & 0.817 & 0.835 & \bf{0.850} \\
			LPIPS$\downarrow$        & 0.510                                                    & 0.523                                            &                                        & 0.516                 & 0.476                     & 0.498 &  & 0.359 & 0.325 & 0.356 & 0.301 & 0.262 & \bf{0.224} \\
			\midrule
			Runtime (s)              & -                                                        & -                                                &                                        & -                     & -                         & -     &  & 0.06  & 0.11  & 0.13  & 0.26  & 0.16  & 0.12       \\
			Params (M)               & -                                                        & -                                                &                                        & -                     & -                         & -     &  & 1.2   & 0.6   & 60.9  & 5.4   & 6.8   & 7.4        \\
			\bottomrule
		\end{tabular}
	}
	\label{tab:lolblur_results}
\end{table*}
\begin{figure*}[t]
	\begin{center}
		\includegraphics[width=\linewidth]{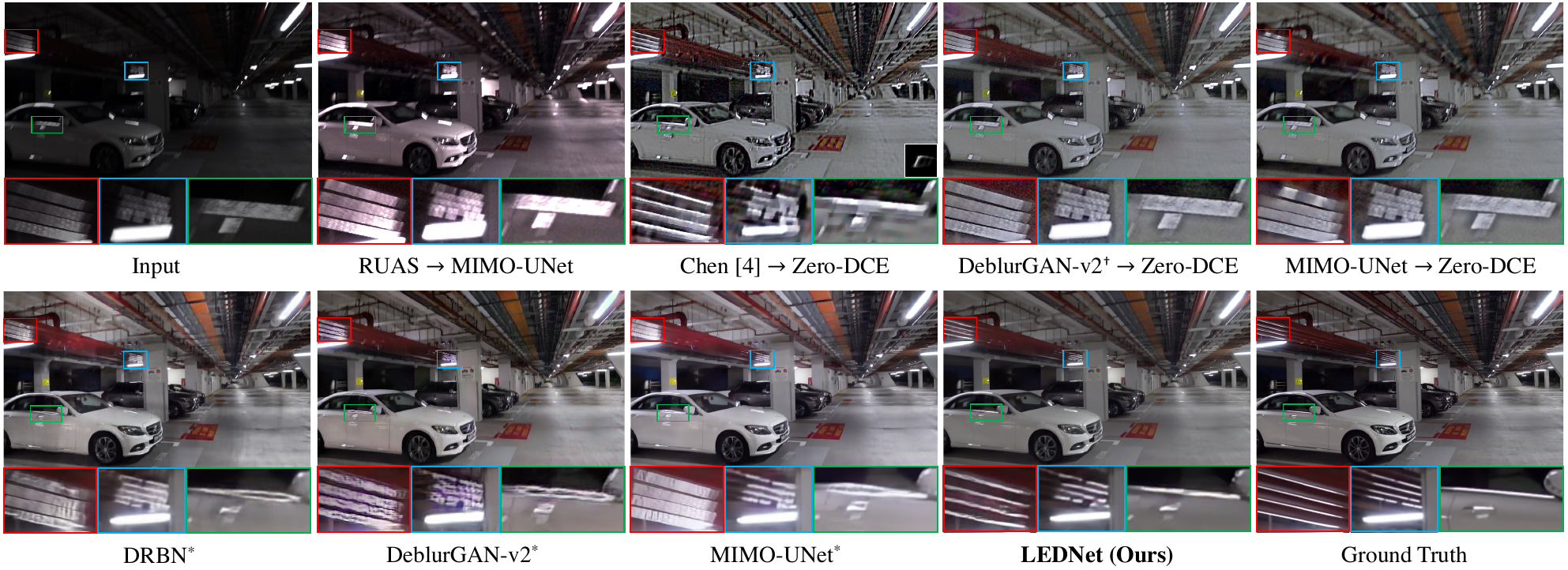}
		\caption{{Visual comparisons on the LOL-Blur dataset. The proposed method generates much sharper images with visually pleasing results. (\textbf{Zoom in for best view})}}
		\label{fig:lolblur_results}
	\end{center}
\end{figure*}
%

\noindent\textbf{Evaluation Metrics.}
We employ the PSNR and SSIM metrics for evaluation on the synthetic LOL-Blur dataset.
To evaluate the perceptual quality of restored images, we also report the perceptual metric LPIPS~\cite{zhang2018unreasonable} for references.
%
%

%
\noindent\textbf{Quantitative Evaluations.}
Table~\ref{tab:lolblur_results} shows quantitative results on our LOL-Blur dataset.
The proposed LEDNet performs favorably against other baseline methods. Notably, the better performance at a similar runtime cost and model size of other networks.
The results suggest LEDNet is effective and particularly well-suited for this task due to the specially designed network and losses.
%
%
%
%

%
\noindent\textbf{Qualitative Evaluations.}
Fig.~\ref{fig:lolblur_results} compares the proposed LEDNet with baseline methods on LOL-Blur dataset.
%
All compared methods produce unpleasing results and suffer from serious blur artifacts, especially in saturated regions.
In contrast, LEDNet generates perceptually pleasant results with sharper textures.
%
%
%
\begin{figure*}[t]
	\begin{center}
		\includegraphics[width=\linewidth]{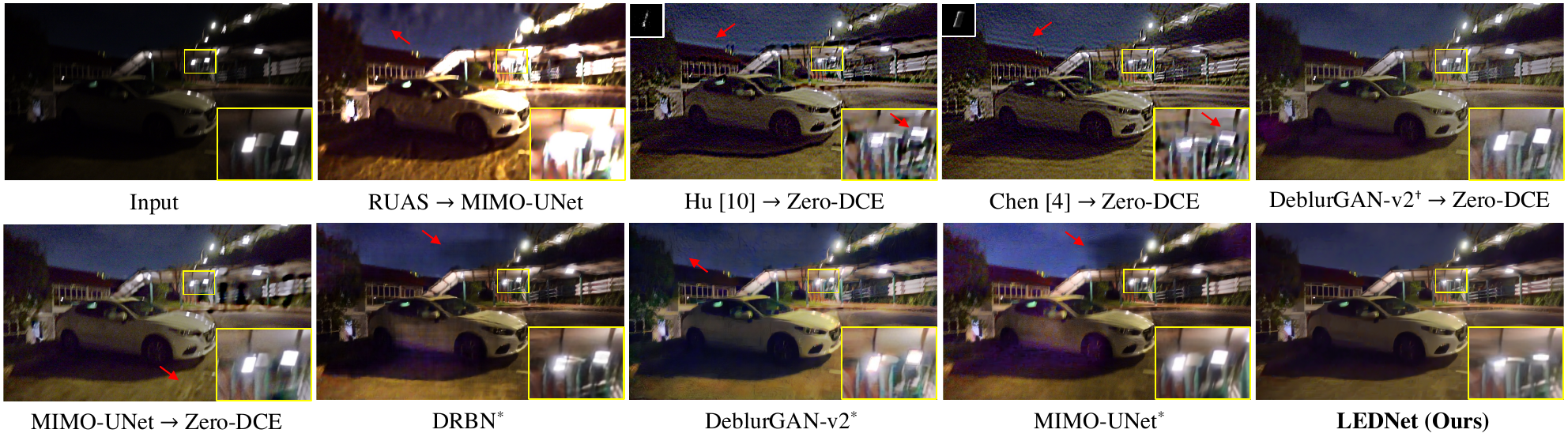}
		\caption{{Visual comparison on a real-world night blurred image. Our method achieves the best perceptual quality with more stable light enhancement and better deblurring performance, while other methods still leave large blurs in saturated regions and suffer from noticeable artifacts, as indicated by red arrows. (\textbf{Zoom in for best view}) }}
		\label{fig:real_results}
	\end{center}
\end{figure*}
\subsection{Evaluation on Real Data}
We also collected a real test dataset, named \textbf{Real-LOL-Blur}, that contains 240 captured low-light blurry images in the wild and 160 night blurry images from the RealBlur dataset~\cite{rim2020real}.
%
%

%
\noindent\textbf{Evaluation Metrics.}
As ground-truths are not available for real test images,
We employ the recent image quality accessment method: MUSIQ~\cite{ke2021musiq}, and two wildely-used ones: NRQM~\cite{ma2017learning} and NIQE~\cite{mittal2012making} as our perceptual metrics.
We choose MUSIQ model trained on KonIQ-10k dataset, it focuses more on color contrast and sharpness assessment, which is more suitable for our task.

\noindent\textbf{Quantitative Evaluations.}
As shown in Table~\ref{tab:real_test}, the proposed LEDNet achieves the highest MUSIQ score, indicating that our results are perceptually best in terms of color contrast and sharpness.
LEDNet also obtains the best NRQM and NIQE scores, showing our results have the best image qualities that are well in line with human perception.

\noindent\textbf{Qualitative Evaluations.}
Fig.~\ref{fig:real_results} presents a visual comparison on a real-world night blurry image.
The methods of Hu~\etal~\cite{hu2014deblurring} and Chen~\etal~\cite{chen2021blind} are particularly designed for saturated image deblurring, however, their cascading baselines still suffer from noticeable artifacts in the presence of large saturated regions.
Besides, the end-to-end baseline networks trained on our LOL-Blur dataset are also less effective given the real-world inputs, as their architecture are not specially designed to handle this task.
As shown in Fig.~\ref{fig:real_results}, their results usually suffer from undesired severe artifacts (red arrows) and blurs (yellow boxes) in their results.
Overall, the proposed LEDNet shows the best visual quality, with fewer artifacts and blurs.
The better performance is attributed to the CurveNLU and FASC, which enable LEDNet to perform spatially-varying feature transformation for both intensity enhancement and blur removal.
The comparisons on real images strongly suggest the effectiveness of our dataset and network.
%
%
Notably, benefiting from the noise simulation via CycleISP in the training dataset, our model handles real-world noises well. 
%

\begin{table}[h]
	\begin{minipage}[t]{0.585\textwidth}
		\centering
		\caption{{Evaluation on Real-LOL-Blur.}}
		\setlength\tabcolsep{2pt}
		\resizebox{1\linewidth}{!}{
			\begin{tabular}{l c c c c c c c}
				\toprule
				                 & RUAS               & MIMO                   & KinD++$^*$ & DRBN$^*$ & DMPHN$^*$ & MIMO$^*$ & Ours       \\
				                 & $\rightarrow$ MIMO & $\rightarrow$ Zero-DCE &            &          &           &          &            \\\midrule
				MUSIQ$\uparrow$  & 34.39              & 28.36                  & 31.74      & 31.27    & 35.08     & 35.37    & \bf{39.11} \\ \midrule
				NRQM$\uparrow$   & 3.322              & 3.697                  & 3.854      & 4.019    & 4.470     & 5.140    & \bf{5.643} \\ \midrule
				NIQE$\downarrow$ & 6.812              & 6.892                  & 7.299      & 7.129    & 5.910     & 4.851    & \bf{4.764} \\ \bottomrule
			\end{tabular}
		}
		\label{tab:real_test}
	\end{minipage}
	\hspace{0.7mm}
	\begin{minipage}[t]{0.4\textwidth}
		\centering
		\caption{{Ablation study results of variant networks on LOL-Blur.}}
		\resizebox{1\linewidth}{!}{
			\begin{tabular}{lccccc}
				\toprule
				               & \begin{tabular}[c]{@{}c@{}}(a)\\ w/o PPM\end{tabular} & \begin{tabular}[c]{@{}c@{}}(b)\\ w/o CurveNLU\end{tabular} & \begin{tabular}[c]{@{}c@{}}(c)\\ Concat\end{tabular} & \begin{tabular}[c]{@{}c@{}}(d)\\ w/o $\mathcal{L}_{en}$\end{tabular} & \begin{tabular}[c]{@{}c@{}}(e)\\ Ours\end{tabular} \\
				\midrule
				PSNR$\uparrow$ & 21.85                         & 25.20                         & 25.31                         & 24.05                         & \bf{25.74 }                   \\ \midrule
				SSIM$\uparrow$ & 0.781                         & 0.823                         & 0.826                         & 0.784                         & \bf{0.850 }                   \\ \bottomrule
			\end{tabular}
		}
		\label{tab:ablation}
	\end{minipage}
\end{table}

\subsection{Ablation Study}
\begin{figure}[t]
	\begin{center}
		\includegraphics[width=\linewidth]{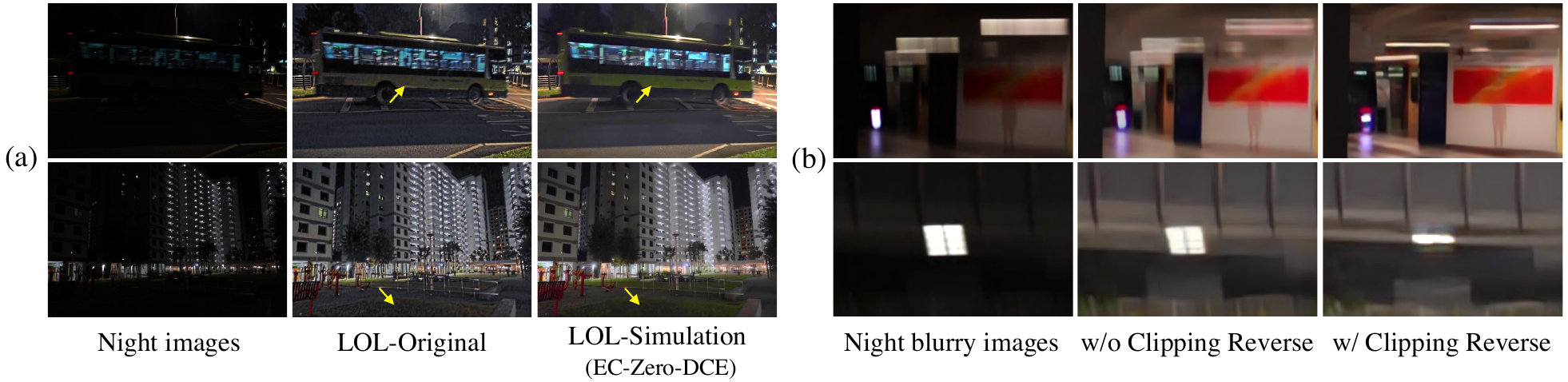}
		\caption{{Ablation study on data synthesis pipeline.
				(a) Results comparison on different training datasets: original LOL dataset (LOL-Original) and our simulated LOL dataset (LOL-Simulation). The network trained on LOL-Simulation generates more natural results with less noise and color distortion.
				(b) Results comparison on different data synthesis pipelines. Applying Clipping Reverse in training data generation enables the network to be robust to handle blur in saturated regions.}}
		\label{fig:blur_simulation}
	\end{center}
\end{figure}

In this subsection, we present an ablation study to demonstrate the effectiveness of the key steps in data synthesis pipeline and the main modules in LEDNet.
\noindent\textbf{Low-light Simulation using EC-Zero-DCE.}
To demonstrate the effectiveness of the proposed low-light simulation,
we construct a new LOL-simulation dataset by applying our EC-Zero-DC to darken the normal-light images in LOL dataset~\cite{wei2018deep}, thus we obtain low-/normal-light paired images for training.
We retrain the network KinD++~\cite{zhang2021beyond} using the LOL-simulation dataset for comparison with the official model that was trained on the original LOL dataset.
Fig.~\ref{fig:blur_simulation}(a) shows that our simulated method enables the network to generate more natural results with less noise and color distortion (indicated by the yellow arrows).

\begin{figure}[tb]
	\centering
	\begin{minipage}[t]{0.565\textwidth}
		\centering
		\includegraphics[width=\linewidth]{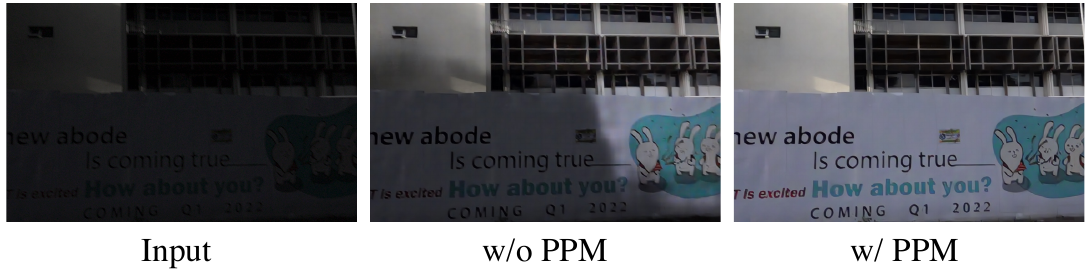}
		\caption{{Result comparison of variant networks: without PPM and with PPM.}}
		\label{fig:ppm}
	\end{minipage}
	\hspace{0.6mm}
	\begin{minipage}[t]{0.4\textwidth}
		\centering
		\includegraphics[width=\linewidth]{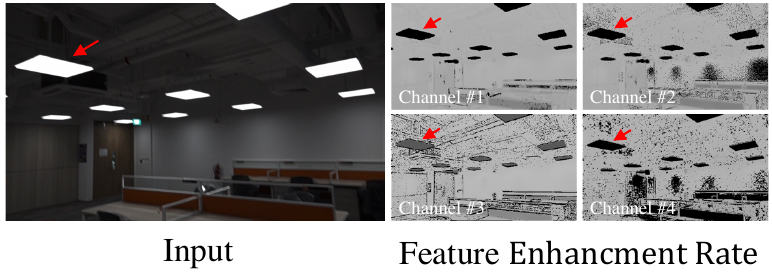}
		\caption{{CurveNLU enhancement rate $F_{out}/F_{in}$ of different channels.
		}}
		\label{fig:enhanced_rate}
	\end{minipage}
\end{figure}

\noindent\textbf{Clipping Reverse (CR).}
Fig.~\ref{fig:data_pipeline}(b) shows that CR helps generate more realistic blurs in saturated regions.
Fig.~\ref{fig:blur_simulation}(b) provides a visual comparison on real-world blurry image, it suggests that applying CR in training data generation helps the network to generalize better in blur removal around saturated regions.
\noindent\textbf{Effectiveness of PPM.}
%
The PPM layer provides crucial global prior for stable training and artifacts suppression in low-light enhancement.
In Table~\ref{tab:ablation}(a), The variant LEDNet without Pyramid Pooling Module (w/o PPM) significantly degrades the network performance.
Besides, the network removing PPM suffers from noticeable artifacts in the enhanced images, as shown in Fig.~\ref{fig:ppm}.
\noindent\textbf{Effectiveness of CurveNLU.}
Fig.~\ref{fig:enhanced_rate} shows the feature enhancement rate $F_{in}/F_{out}$ of input $F_{in}$ and output $F_{out}$ of CurveNLU.
%
%
As observed, feature adjustment in CurveNLU is spatially adaptive to different regions in the image.
The merit of CurveNLU is also validated in Table~\ref{tab:ablation}.
\noindent\textbf{Effectiveness of FASC Connections.}
Comparing variant LEDNet (c) and (e) in Table~\ref{tab:ablation}, the one with FASC connection achieves better performance compared to simple connection based on concatenation.
This is because the saturated and unsaturated areas in the night scenes follow different blur models. The task in this paper poses more requirements of spatially-varying operations.
\noindent\textbf{Effectiveness of Enhancement Loss.}
The enhancement loss $\mathcal{L}_{en}$ is necessary in our method.
Removing it from training harm the performance as shown in Table~\ref{tab:ablation}(d).
It is because this intermediate loss helps decompose our joint task into low-light enhancement and deblurring, which makes it easier to optimize.

\section{Conclusion}
We have presented a novel data synthesis pipeline to model realistic low-light blurring.
Based on the pipeline, we built a large-scale and diverse paired dataset (LOL-Blur) for learning and benchmarking the new joint task of low-light enhancement and deblurring.
We have also proposed a simple yet effective model, LEDNet, which performs illumination enhancement and blur removal in a single forward pass.
We showed that PPM is beneficial and introduced CurveNLU to make the learned network more stable and robust. We further described FASC for better deblurring.
Our dataset and network offer a foundation for further exploration for low-light enhancement and deblurring in the dark.
%
%

\vspace{1mm}
\noindent {\bf Acknowledgement.}
%
This study is supported under the RIE2020 Industry Alignment Fund – Industry Collaboration Projects (IAF-ICP) Funding Initiative, as well as cash and in-kind contribution from the industry partner(s).
%
\clearpage

\bibliographystyle{splncs04}
\bibliography{led_bib}

\clearpage
\renewcommand\thesection{\Alph{section}}
\setcounter{section}{0}
\begin{center}
	{\quad}\\
	\vspace{20pt}
	\Large\textbf{{LEDNet: Joint Low-light Enhancement and Deblurring in the Dark}}\\
	\vspace{15pt}
	\Large{Supplementary Material} \\
	\vspace{25pt}
\end{center}

In this supplementary material, we provide additional details and results to the paper. In Sec.~\ref{sec:arch}, we first present the architecture details of some modules in our proposed LEDNet.
Sec.~\ref{sec:lednet} presents further analysis and discussions on our proposed LEDNet network, consisting of more analysis and results on CurveNLU, and loss function.
In Sec.~\ref{sec:lolblur}, we provide more analysis on our data synthesis pipeline and LOL-Blur dataset.
Finally, more visual comparisons on both LOL-Blur  and real-world images are provided in Sec.~\ref{sec:results}.
In addition, we present a video demo to show the effectiveness of proposed LEDNet for dealing with real-world dark blurry videos.
\vspace{4mm}
\section{Architecture Details}
\label{sec:arch}
\begin{figure*}[h]
	\centering
	\vspace{-3mm}
	\includegraphics[width=0.99\textwidth]{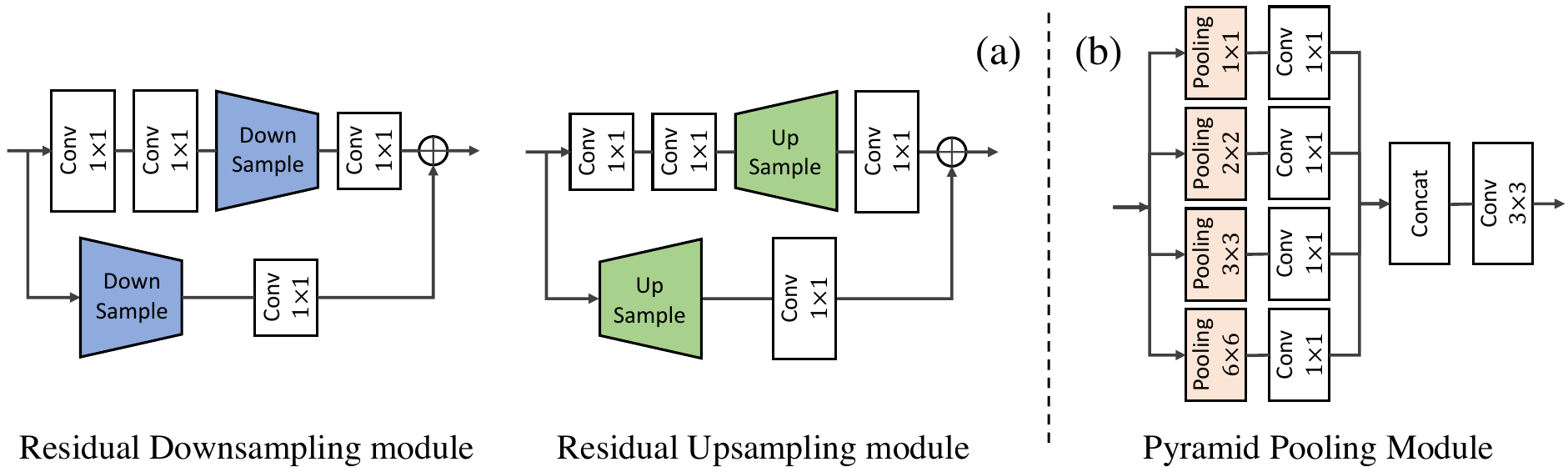}
	\caption{\footnotesize{Illustration of the Residual Downsample/Upsample Modules~\cite{zamir2020learning} and the Pyramid Pooling Module~\cite{zhao2017pyramid}.
		}}
	\label{fig:modules}
	\vspace{-2mm}
\end{figure*}

As shown in Fig. 4 in the main manuscript, we adopt the existing Residual Downsample/Upsample~\cite{zamir2020learning} and Pyramid Pooling Module (PPM)~\cite{zhao2017pyramid} in our LEDNet.
For reading convenience, Fig.~\ref{fig:modules}(a) and (b) provide the detailed structures of Residual Downsample/Upsample and PPM, which are the same as their original configurations.

\section{More Discussions on LEDNet}
\label{sec:lednet}
In this section, we first present more ablation experiments to show the effect of the key components of the proposed LEDNet, including CurveNLU, PPM, and enhancement loss.

\subsection{Analysis on CurveNLU}
\begin{figure*}[h]
	\centering
	\vspace{1mm}
	\includegraphics[width=0.98\textwidth]{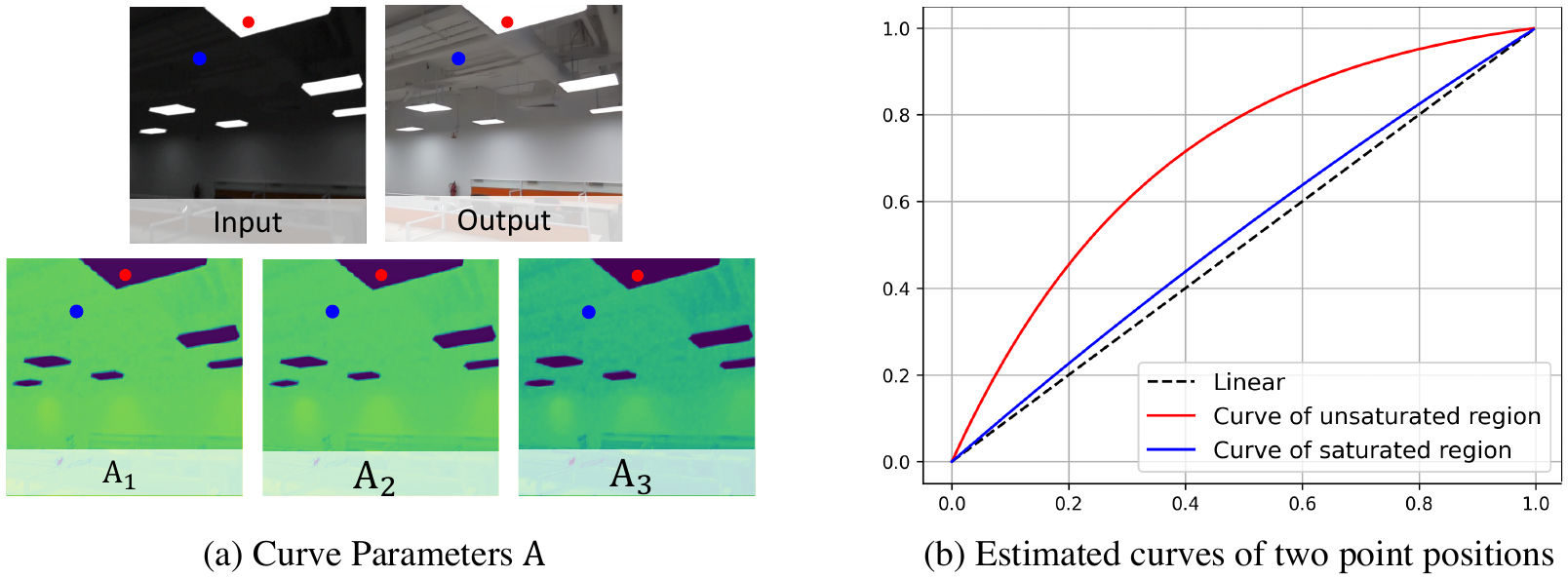}
	\caption{\footnotesize{(a) Visualization of  estimated curve parameters. (b) Estimated curves of two points in the unsaturated region and saturated region.
		}}
	\label{fig:curve_param}
\end{figure*}

\vspace{1mm}
\noindent\textbf{Curve Parameter Visualization.}
To further explore CurveNLU, we visualize an example of estimated curve parameters  $A$ in Fig.~\ref{fig:curve_param}(a).
The parameters are significantly different between unsaturated regions and saturated regions.
Fig.~\ref{fig:curve_param}(b) shows two estimated curves of blue and red points in Fig.~\ref{fig:curve_param}(a), which lay in the unsaturated region and saturated region respectively.
The red curve of the unsaturated region has the greater curvature, thus, there is a lager feature intensity increase for darkder areas.
In contrast, the blue curve of the saturated region that has a curvature close to 0 tends to maintain the feature value in the saturated regions.
Therefore, the non-linear CurveNLU modules can increase intensity for dark areas while avoiding overexposure in the saturation regions.
\noindent\textbf{Effect of  Curve Order.}
To explore the effect of curve order $n$ in the CurveNLU module, we conduct experiments that use different $n$  for comparison.
As shown in Table~\ref{tab:n_curenlu}, using the higher curve order $n$ over 3 only leads to slight PSNR/SSIM gains. Thus we use $n = 3$ in our CurveNLU modules as a trade-off between computational complexity and performance.
Notably, compared with the baseline without CurveNLU inserted, i.e., $n = 0$, our proposed LEDNet obtains a large performance gain.

\begin{table}[h]
	\caption{\footnotesize{Results on LOL-Blur dataset for different curve orders in CurveNLU modules. }}
	\footnotesize
	\centering
	\vspace{2mm}
	\smallskip
	\setlength\tabcolsep{4pt}
	\resizebox{0.77\textwidth}{!}{
		\begin{tabular}{c c cc c c}
			\toprule
			     & $n$ = 0 (w/o CurveNLU) & $n$ = 1 & $n$ = 2 & $n$ = 3 (Ours) & $n$ = 4 \\
			\midrule
			PSNR & 25.20                  & 25.25   & 25.48   & 25.74          & 25.77   \\
			SSIM & 0.823                  & 0.826   & 0.838   & 0.850          & 0.850   \\
			\bottomrule
		\end{tabular}
	}
	\label{tab:n_curenlu}
\end{table}
\subsection{Effectiveness of Enhancement Loss}
\begin{figure*}[h]
	\centering
	\vspace{0mm}
	\includegraphics[width=0.99\textwidth]{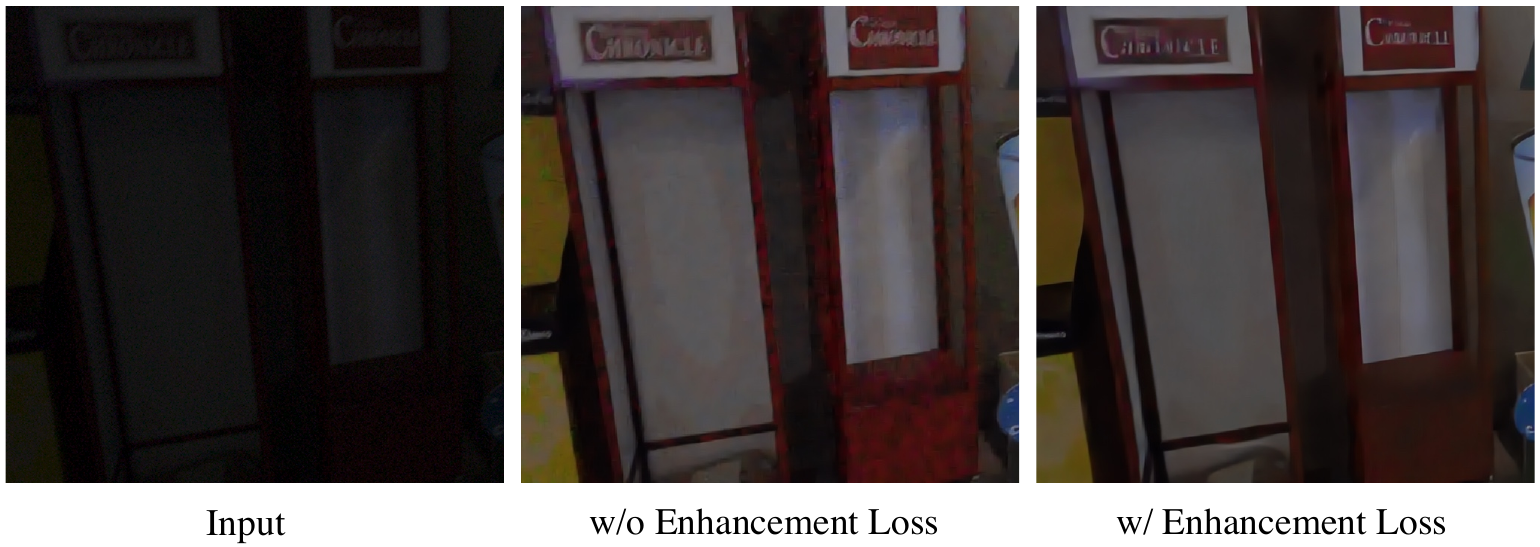}
	\caption{\footnotesize{Visual comparison of different losses that using enhancement loss or not.}}
	\label{fig:enhance_loss}
\end{figure*}

Table 2 in the main manuscript has suggested that using enhancement loss $\mathcal{L}_{en}$ is indispensable in our method.
Fig.~\ref{fig:enhance_loss} further shows removing the $\mathcal{L}_{en}$ in the training process harms the visual quality significantly.
The network trained without $\mathcal{L}_{en}$ produces severe artifacts with unsmooth regions in the result.
%
\section{More Discussions on LOL-Blur Dataset}
\label{sec:lolblur}
%

\subsection{Simulation of Low light}
\begin{figure*}[h]
	\centering
	\includegraphics[width=0.99\textwidth]{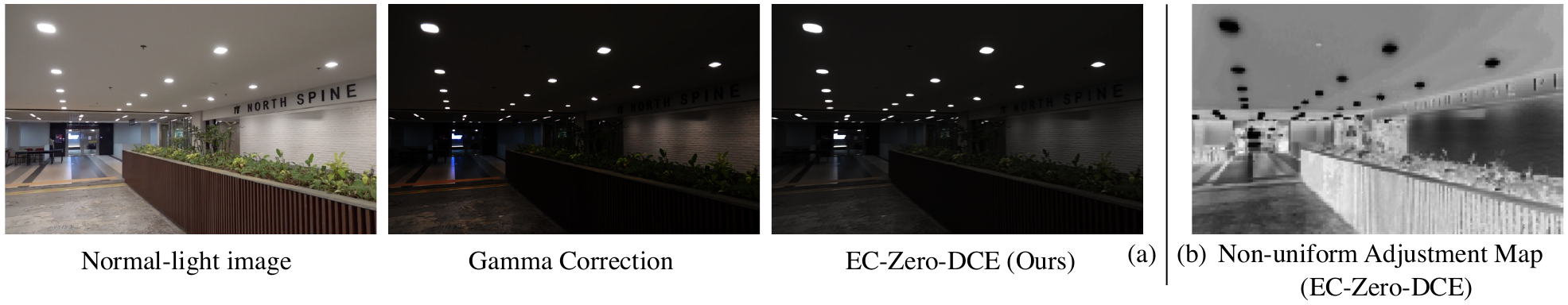}
	\vspace{-1mm}
	\caption{\footnotesize{(a) Comparisons between the proposed EC-Zero-DCE and Gamma correction for generating low-light images.
			(b) The non-uniform adjustment map suggests that CE-Zero-DCE performs spatially-varying luminance adjustment.}}
	\label{fig:low_light}
	\vspace{-2mm}
\end{figure*}

\begin{figure*}[h]
	\begin{center}
		\includegraphics[width=0.99\linewidth]{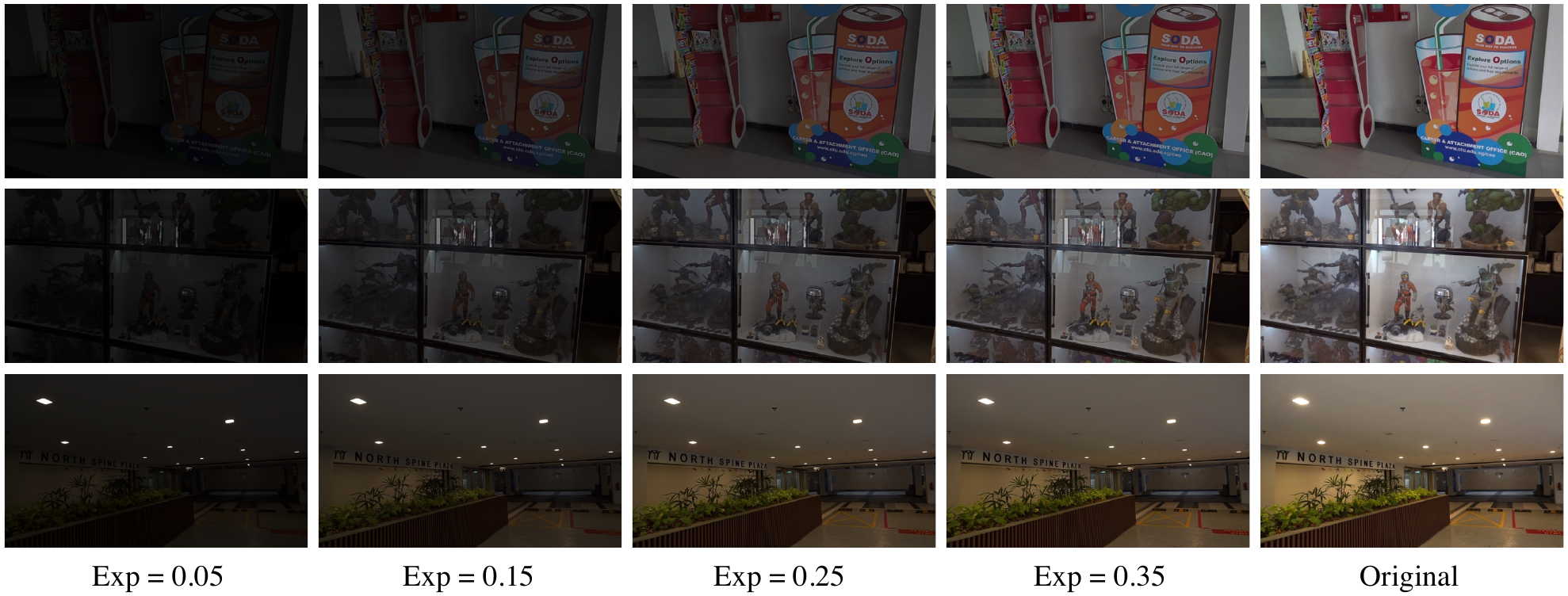}
		\vspace{-1mm}
		\caption{\footnotesize{The proposed EC-Zero-DCE is able to generate low-light images with different exposure levels, while keeping the intensities of saturated pixels unchanged. }}
		\label{fig:exposure_levels}
	\end{center}
	\vspace{-4mm}
\end{figure*}

In this paper, we use the Exposure Conditioned Zero-DCE (EC-Zero-DCE) to generate the low-light images of different exposure levels.
Fig.~\ref{fig:low_light}(a) compares our low-light data synthesis pipeline with Gamma correction that has been used in prior works~\cite{lore2017llnet,lv2018mbllen}.
As we can see from this comparison, the image generated by Gamma correction has a large color deviation with noticeable warm tones.
In contrast, our EC-Zero-DCE can produce more natural and realistic low-light images.
Moreover, the proposed EC-Zero-DCE performs pixel-wise and spatially-varying light adjustment, Fig.~\ref{fig:low_light}(b) provides a non-uniform luminance adjustment map for this case.
Given different exposure levels, EC-Zero-DCE can generate realistic low-light images with diverse darkness, as shown in Fig.~\ref{fig:exposure_levels}.
Thanks to its spatially-varying adjustment, EC-Zero-DCE tends to retain the intensities of saturated pixels, simulating the realistic light effect in real-world night images.

\subsection{Simulation of Noise}
\begin{figure*}[h]
	\centering
	\includegraphics[width=0.99\textwidth]{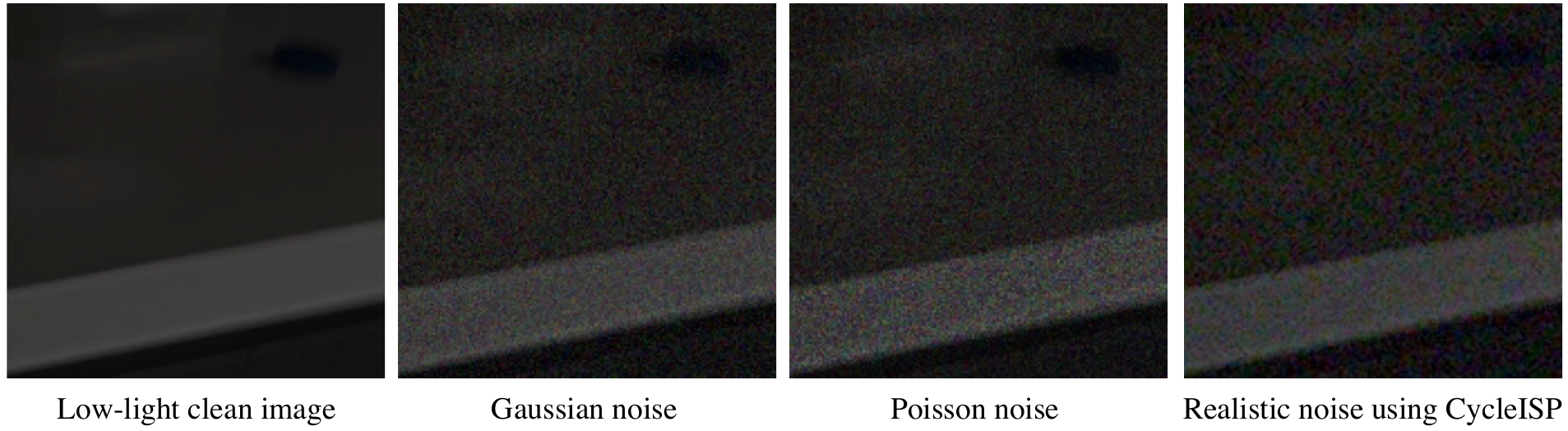}
	\caption{\footnotesize{Noise simulation comparison among CycleISP, Gaussian, and Poisson.}}
	\label{fig:noise}
	\vspace{-2mm}
\end{figure*}

To simulate realistic noise in dark images, we adopt CycleISP~\cite{zamir2020cycleisp} to generate the noisy image in the RAW domain.
We compare our noise simulation with Gaussian and Poisson noise that are commonly used in other restoration tasks, e.g., blind face restoration~\cite{li2020blind,wang2021towards} and real-world blind super-resolution~\cite{wang2021realesrgan,zhang2021designing,chan2021realbasicvsr}.
Fig.~\ref{fig:noise} shows the noises generated by CycleISP are more natural and realistic.
%
\subsection{Luminance Distribution of Datasets}
\begin{figure*}[h]
	\centering
	\includegraphics[width=0.99\textwidth]{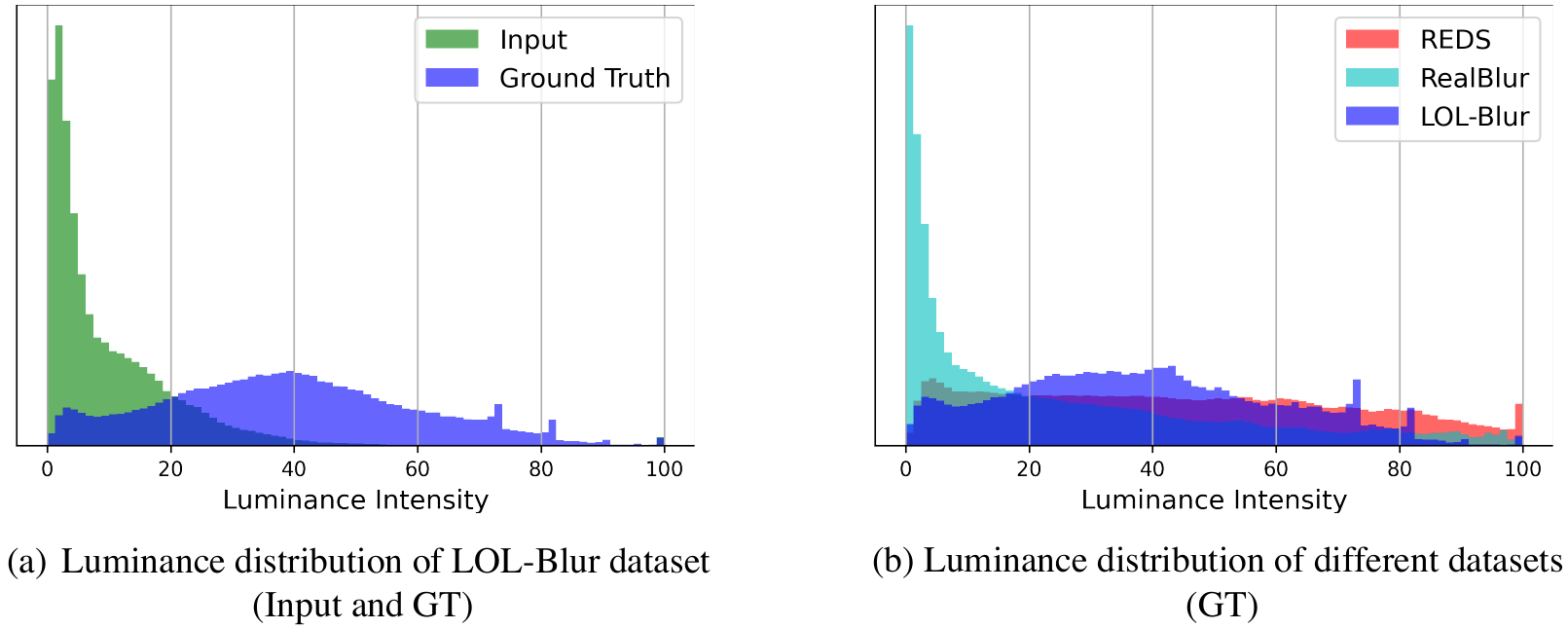}
	\caption{\footnotesize{(a) Luminance distribution of LOL-Blur dataset. (b) Comparison of Luminance distributions of different deblurring datasets.}}
	\label{fig:dataset_dist}
	\vspace{-2mm}
\end{figure*}

Fig.~\ref{fig:dataset_dist}(a) shows the luminance distribution of our proposed LOL-Blur Dataset.
Fig.~\ref{fig:dataset_dist}(b) provides a comparison of luminance distributions of different deblurring datasets.
The great majority brightness of ground truth images in the RealBlur dataset lay the range of small intensity, thus,  RealBlur is not suitable for training a light enhancement network.
Besides, there are many sunny scenes in the REDS dataset, which can not be adopted to generate low-light images.
\section{Qualitative Comparisons}
\label{sec:results}
%
In this section, we present more visual comparisons with the baselines studied in the main manuscript:
RUAS~\cite{liu2021retinex} $\rightarrow$ MIMO-UNet~\cite{cho2021rethinking},
DeblurGAN-v2$^\dagger$~\cite{kupyn2019deblurgan} $\rightarrow$ Zero-DCE~\cite{zerodce},
MIMO-UNet~\cite{cho2021rethinking} $\rightarrow$ Zero-DCE~\cite{zerodce},
DRBN$^*$ ~\cite{yang2020fidelity},
DeblurGAN-v2$^*$~\cite{kupyn2019deblurgan},
DMPHN$^*$~\cite{zhang2019deep},
and MIMO-UNet$^*$~\cite{cho2021rethinking}.
Fig.~\ref{fig:lolblur_results_supp_2} provides more visual comparisons on our LOL-Blur Dataset.
In addition, Figs.~\ref{fig:real_results_supp_1},~\ref{fig:real_results_supp_2}, and~\ref{fig:real_results_supp_3} provide more visual comparisons on real-world night blurry images.
To demonstrate the generalizability in the wild of our dataset and network, we also test on more real-world night blurry images in the RealBlur dataset~\cite{rim2020real}. Fig.~\ref{fig:different_blurs} shows our LEDNet is able to handle various blur patterns (revealed by light streaks in the input images). Besides, Fig.~\ref{fig:real_results_realblur_dataset} provides more results in different scenarios on the RealBlur dataset. 
The above extensive real-world results\footnote{Note that all the {Figs.~\ref{fig:real_results_supp_1},~\ref{fig:real_results_supp_2},~\ref{fig:real_results_supp_3},~\ref{fig:different_blurs}}, and {Fig.~\ref{fig:real_results_realblur_dataset}} provided in this suppl., as well as all results in \href{https://youtu.be/450dkE-fOMY}{[\color{blue}\underline{video demo}}] are evaluated on the real-world scenarios.} suggest that our method performs well on diverse test images and videos in the wild.
%
%
\begin{figure*}[h]
	\centering
	\vspace{2mm}
	\includegraphics[width=0.994\textwidth]{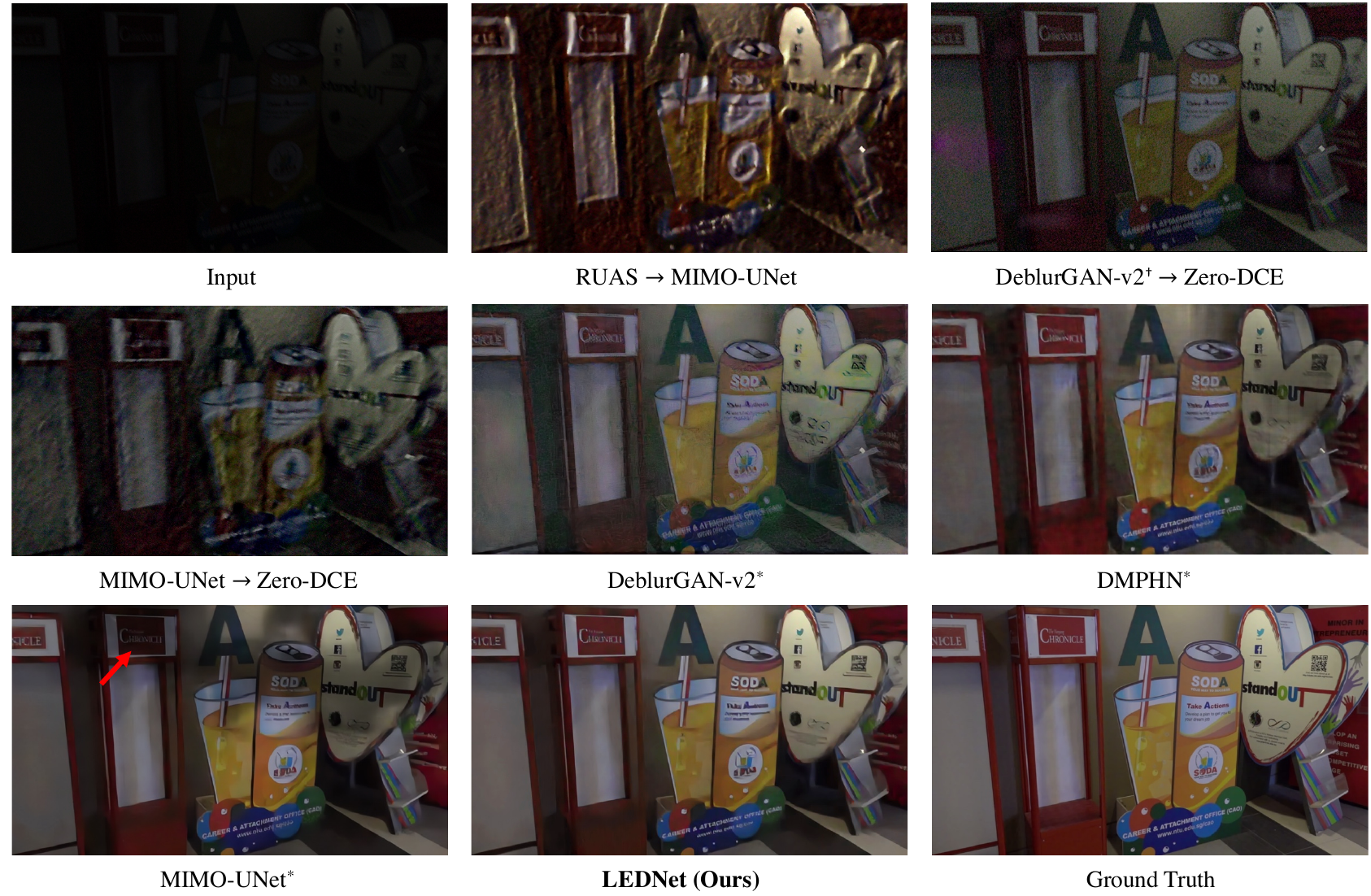}
	\vspace{-6mm}
\end{figure*}
\begin{figure*}[ht]
	\centering
	\includegraphics[width=0.99\textwidth]{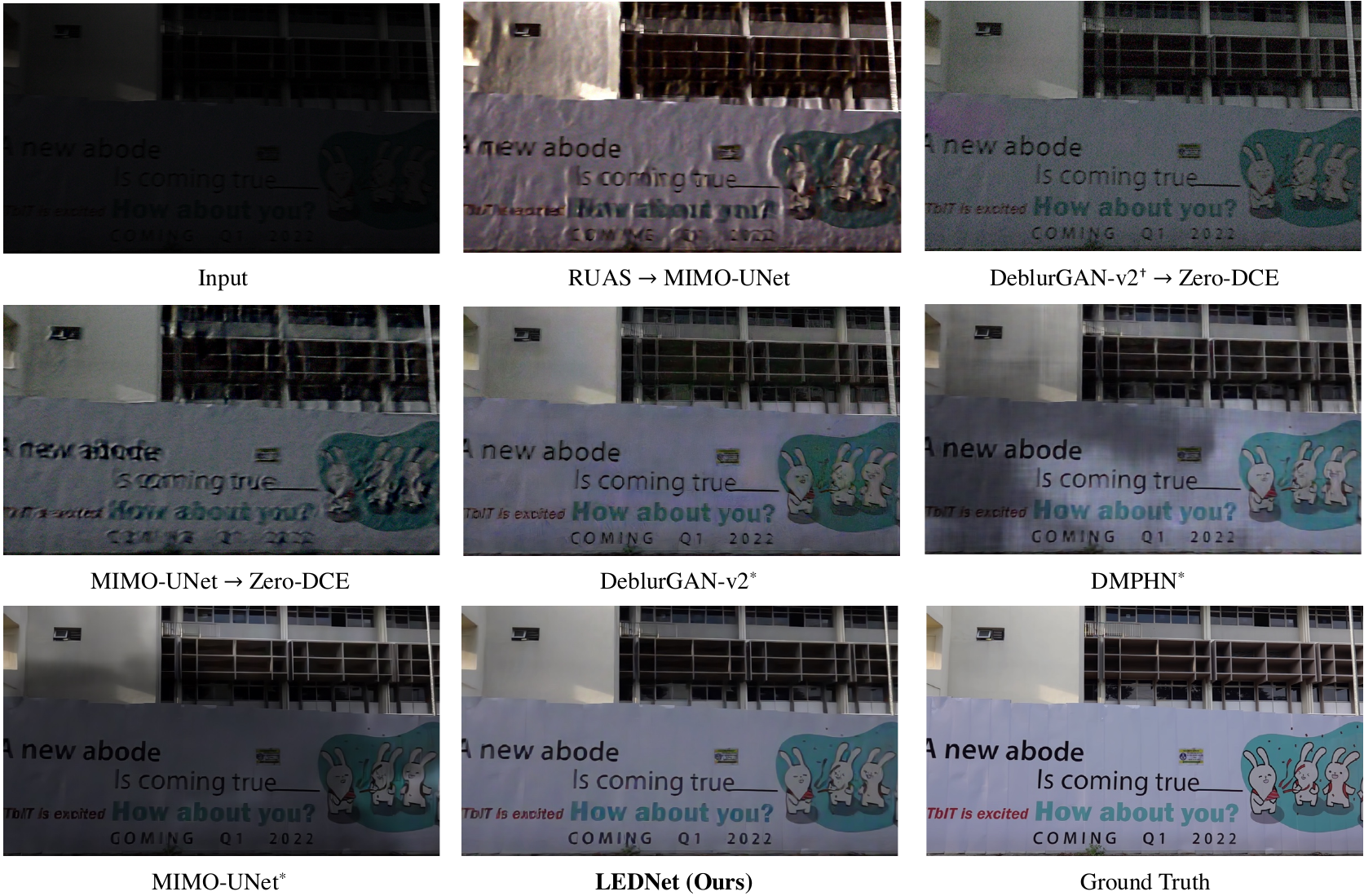}
	\caption{\footnotesize{Visual comparison on our LOL-Blur dataset. The proposed LEDNet generates much sharper images with visually pleasing results.
			The symbol `$^\dagger$' indicates that we use DeblurGAN-v2 trained on RealBlur dataset, and `$^*$' indicates the network is trained with our LOL-Blur dataset. \textbf{(Zoom in for best view)}}}
	\label{fig:lolblur_results_supp_2}
\end{figure*}
%
%
\begin{figure*}[h]
	\centering
	\vspace{-2mm}
	\includegraphics[width=0.99\textwidth]{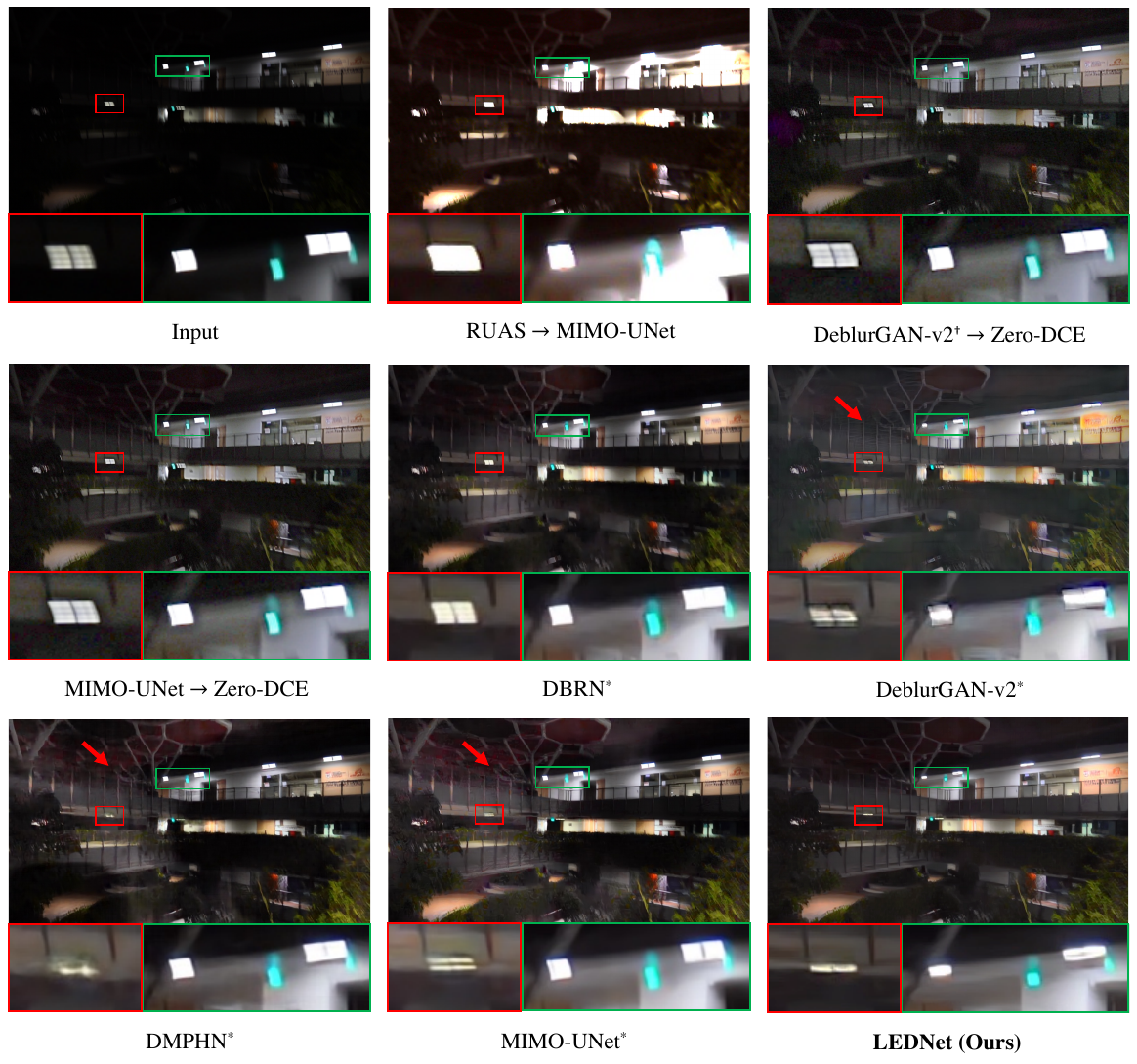}
	\caption{\footnotesize{Visual comparison on a real-world night blurry image. The proposed LEDNet achieves the best perceptual quality with more stable light enhancement and better deblurring performance, especially in saturated regions,
			while other methods still leave large blurs in saturated regions and suffer from noticeable artifacts, as indicated by red arrows.
			The symbol `$^\dagger$' indicates that we use DeblurGAN-v2 trained on RealBlur dataset, and `$^*$' indicates the network is trained with our LOL-Blur dataset. \textbf{(Zoom in for best view)}}}
	\label{fig:real_results_supp_1}
\end{figure*}
\begin{figure*}[h]
	\centering
	\vspace{-2mm}
	\includegraphics[width=0.99\textwidth]{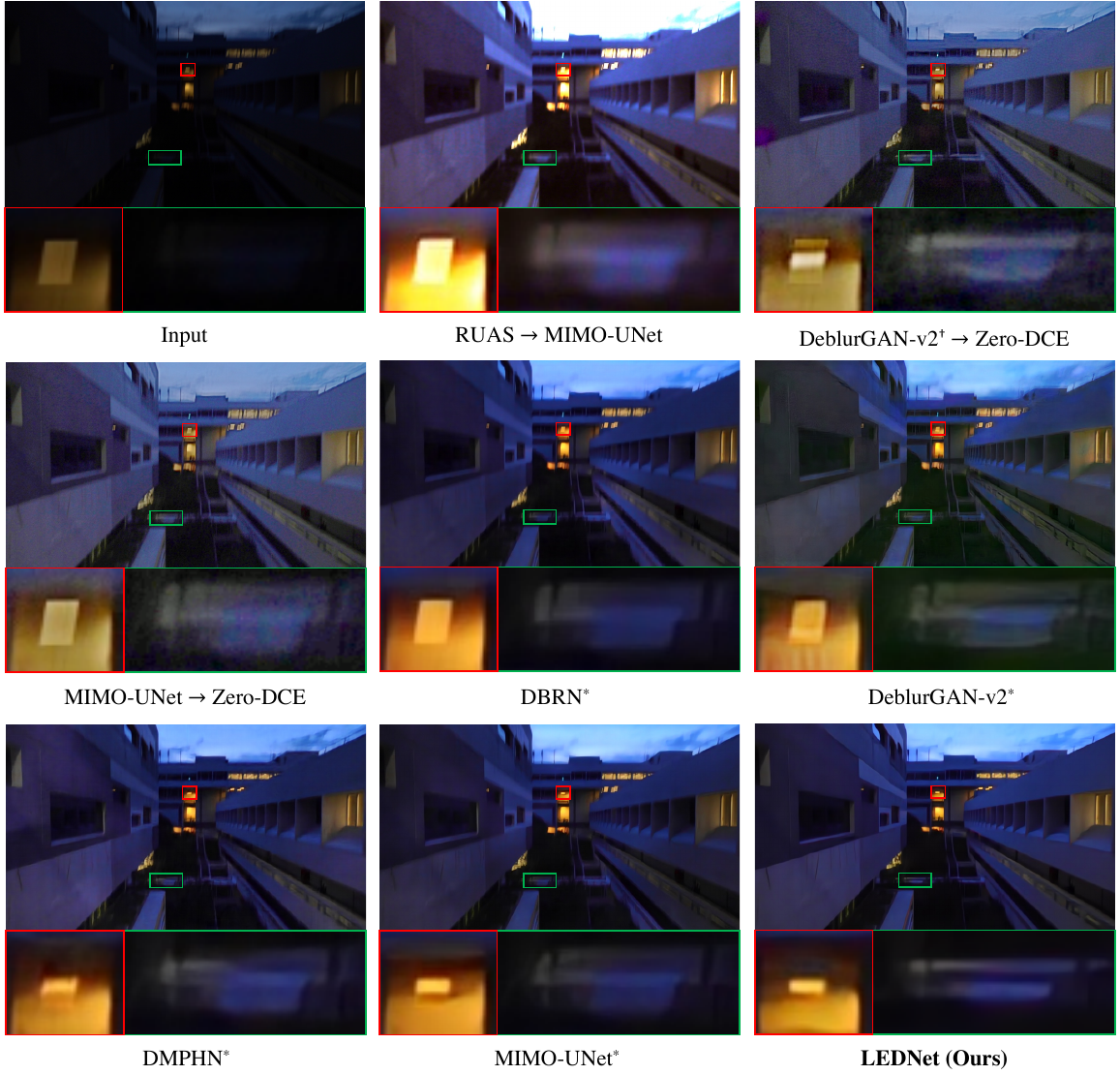}
	\caption{\footnotesize{Visual comparison on a real-world night blurry image. The proposed LEDNet achieves the best perceptual quality with more stable light enhancement and better deblurring performance, especially in saturated regions,
			while other methods still leave large blurs in saturated regions and suffer from noticeable artifacts.
			The symbol `$^\dagger$' indicates that we use DeblurGAN-v2 trained on RealBlur dataset, and `$^*$' indicates the network is trained with our LOL-Blur dataset. \textbf{(Zoom in for best view)}}}
	\label{fig:real_results_supp_2}
\end{figure*}
\begin{figure*}[h]
	\centering
	\vspace{-2mm}
	\includegraphics[width=0.99\textwidth]{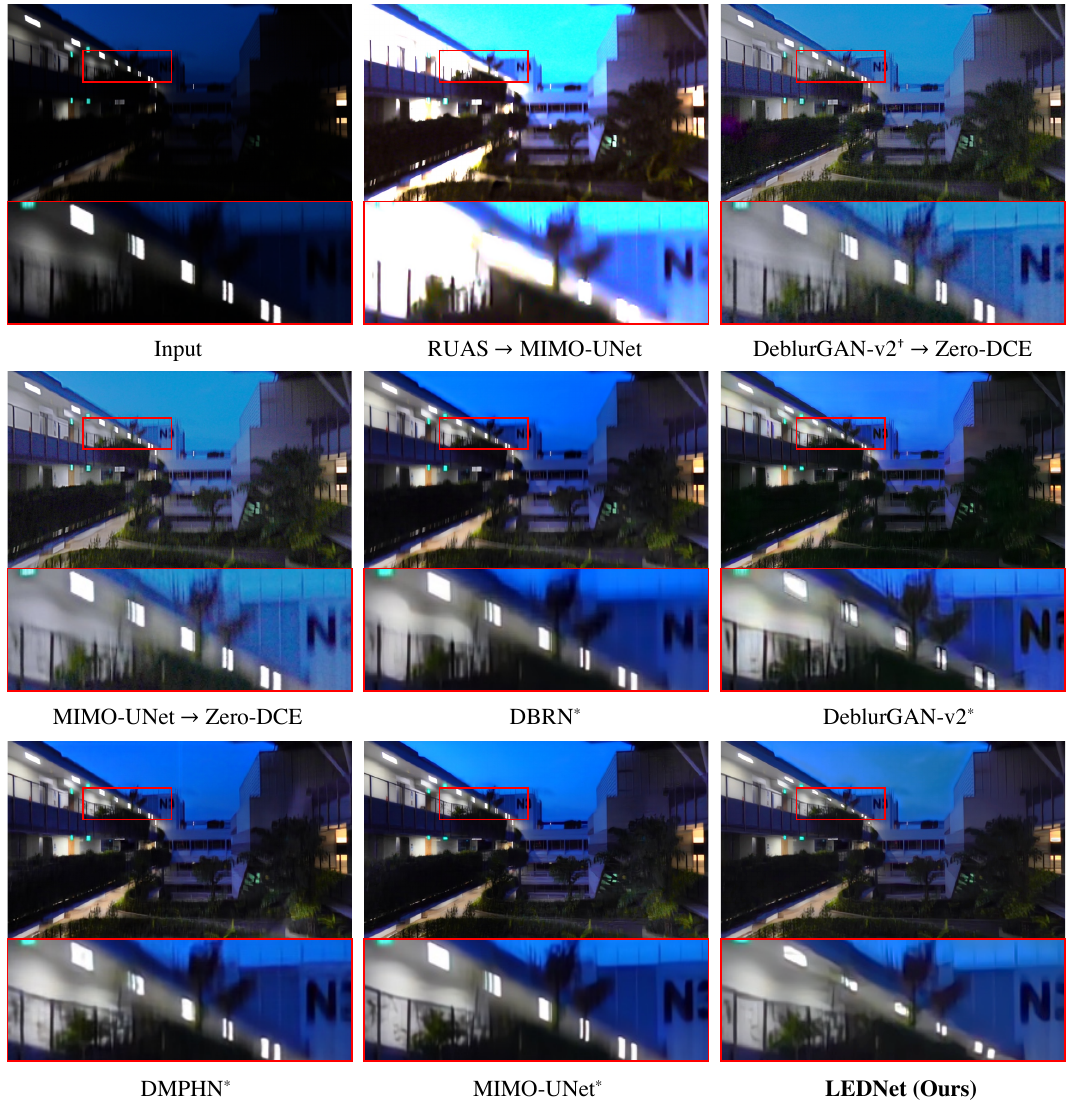}
	\caption{\footnotesize{Visual comparison on a real-world night blurry image. The proposed LEDNet achieves the best perceptual quality with more stable light enhancement and better deblurring performance, especially in saturated regions,
			while other methods still leave large blurs in saturated regions and suffer from noticeable artifacts.
			The symbol `$^\dagger$' indicates that we use DeblurGAN-v2 trained on RealBlur dataset, and `$^*$' indicates the network is trained with our LOL-Blur dataset. \textbf{(Zoom in for best view)}}}
	\label{fig:real_results_supp_3}
\end{figure*}
\begin{figure*}[h]
	\centering
	\vspace{-2mm}
	\includegraphics[width=0.99\textwidth]{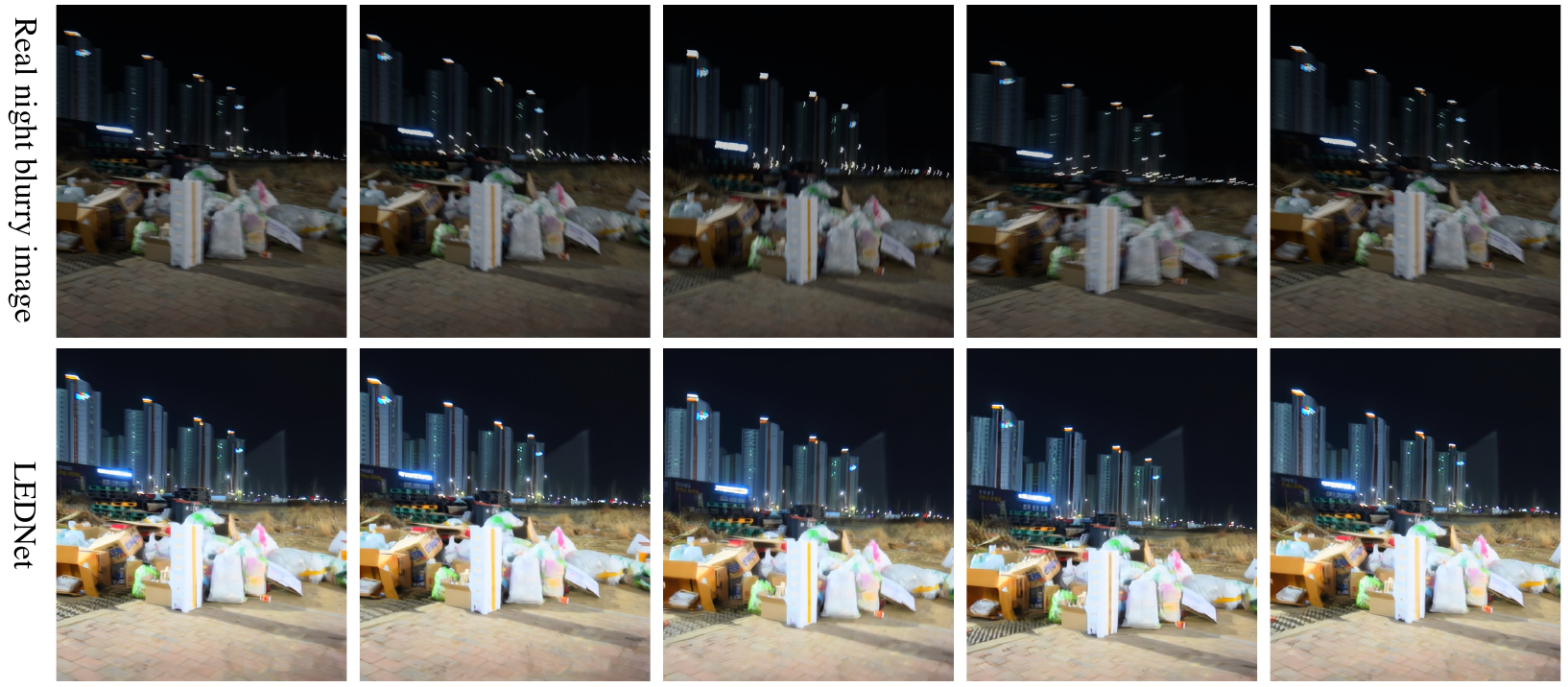}
	\caption{\footnotesize{Evaluation on different blur patterns on RealBlur dataset~\cite{rim2020real}. The proposed LEDNet is able to handle blurs of different shapes, which can be observed from the light streaks in the input images. \textbf{(Zoom in for best view)}}}
	\label{fig:different_blurs}
\end{figure*}
\begin{figure*}[h]
	\centering
	\vspace{0mm}
	\includegraphics[width=0.84\textwidth]{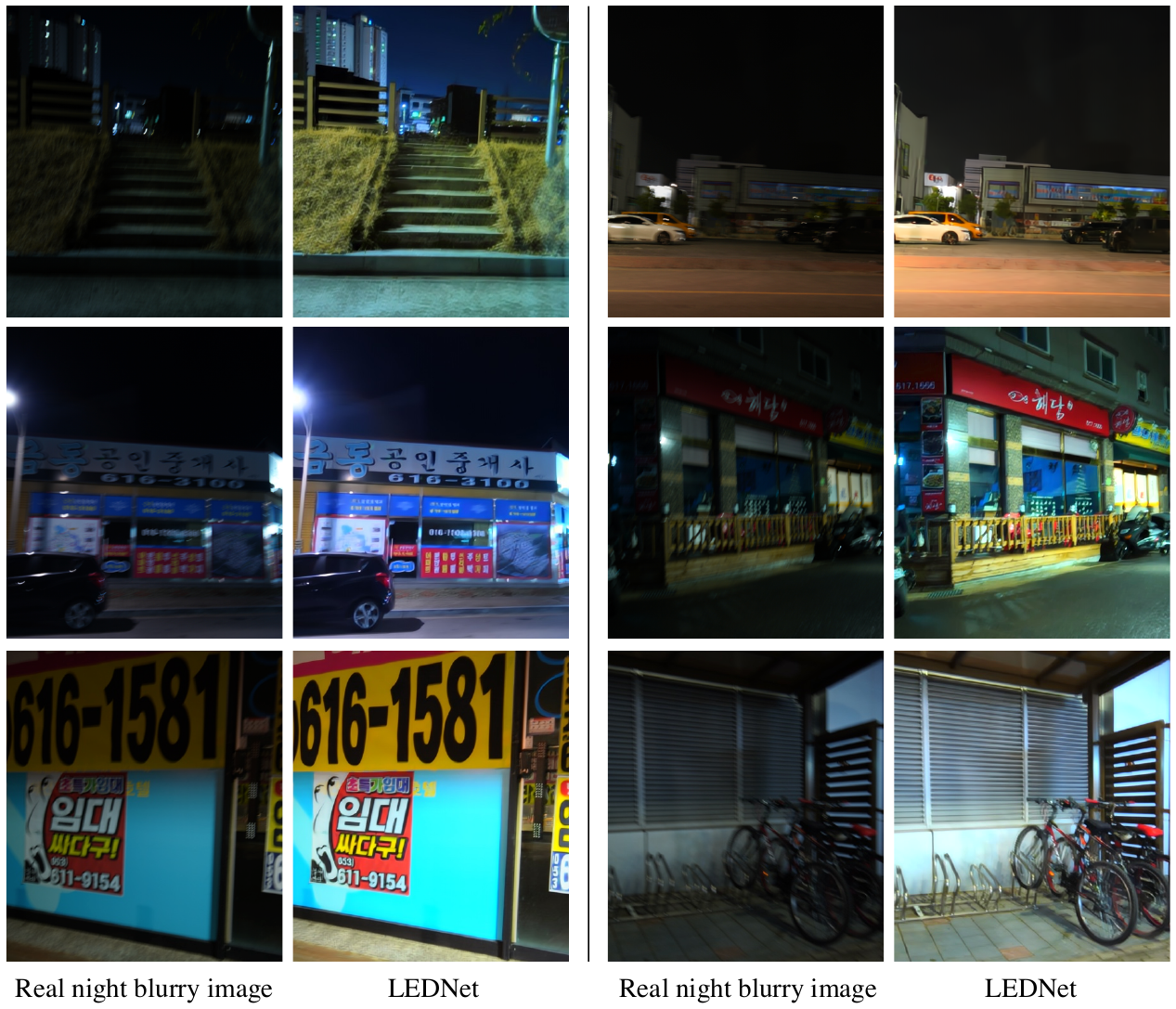}
	\caption{\footnotesize{Visual results on RealBlur dataset~\cite{rim2020real}. The proposed LEDNet performs well in the different scenarios. \textbf{(Zoom in for best view)}}}
	\label{fig:real_results_realblur_dataset}
\end{figure*}

\end{document}